\documentclass[manuscript,screen]{acmart}
\AtBeginDocument{%
  }

\setcopyright{acmlicensed}
\copyrightyear{2018}
\acmYear{2018}
\acmDOI{XXXXXXX.XXXXXXX}
\acmISBN{978-1-4503-XXXX-X/2018/06}

\usepackage{amsmath,amsfonts}
\usepackage{algorithmic}
\usepackage{graphicx}
\usepackage{textcomp}
\usepackage{xcolor}
\usepackage{color}
\usepackage{algorithm,algorithmic}
\usepackage{multirow}
\usepackage{array}
\usepackage{enumitem}
\usepackage{makecell}

\newtheorem*{example}{Example}

\usepackage[compat=0.4]{yquant}
\yquantdefinebox{null}[inner sep=0pt]{}
\usepackage{tikz}
\usepackage{pgfplots}
\pgfplotsset{compat=1.18}
\usetikzlibrary{fit}
\usetikzlibrary{calc,patterns}
\usetikzlibrary{shapes.geometric, arrows, positioning}

\tikzstyle{startstop} = [rectangle, rounded corners, minimum width=1cm, minimum height=1cm, text centered, draw=black]
\tikzstyle{process} = [rectangle, minimum width=1cm, minimum height=1cm, text centered, draw=black]
\tikzstyle{decision} = [diamond, minimum width=1cm, minimum height=0.8cm, text centered, draw=black]
\tikzstyle{arrow} = [thick,->,>=stealth]
\definecolor{darkorange}{rgb}{1.0, 0.55, 0.0} 
\tikzset{
    SSRStyle/.style={color=black, thick},
    QiskitStyle/.style={color=darkorange, thick},
    MCTSStyle/.style={color=green!50!black, thick}
}
\usepackage{subcaption}
\useyquantlanguage{groups}

\begin{document}

\title{SSR: A Swapping-Sweeping-and-Rewriting Optimizer for Quantum Circuit Transformation}


\author{Yunqi Huang}
\affiliation{%
  \institution{ College of Artificial Intelligence, Nanjing Tech University}
  \city{Nanjing}
  \country{China}}
\affiliation{%
  \institution{Centre for Quantum Software and Information (QSI), Faculty of Engineering and Information Technology, University of Technology Sydney}
  \city{Sydney}
  \country{Australia}}

\author{Xiangzhen Zhou}
\correspondingauthor
\affiliation{
 \institution{College of Artificial Intelligence, Nanjing Tech University}
 \city{Nanjing}
 \country{China}
}
\affiliation{
 \institution{Arclight Quantum Computing Inc.}
 \city{Beijing}
 \country{China}
}
\email{xiangzhenzhou@njtech.edu.cn}

\author{Fanxu Meng}
\affiliation{
 \institution{College of Artificial Intelligence, Nanjing Tech University}
 \city{Nanjing}
 \country{China} 
}

\author{Pengcheng Zhu}
\correspondingauthor
\affiliation{
\institution{College of Information Engineering, Taizhou University}
\city{Taizhou}
\country{China}
}
\email{zhupcnt@tzu.edu.cn}

\author{Yu Luo}
\affiliation{
\institution{School of Artificial Intelligence and Computer Science, Shaanxi Normal University}
\city{Xi'an}
\country{China}
}

\author{Zhenlong Du}
\affiliation{
\institution{College of Artificial Intelligence, Nanjing Tech University}
 \city{Nanjing}
 \country{China}
}


\begin{abstract}
Quantum circuit transformation (QCT), necessary for adapting any quantum circuit to the qubit connectivity constraints of the NISQ device, often introduces numerous additional SWAP gates into the original circuit, increasing the circuit depth and thus reducing the success rate of computation. To minimize the depth of QCT circuits, we propose a Swapping-Sweeping-and-Rewriting optimizer. This optimizer rearranges the circuit based on generalized gate commutation rules via a genetic algorithm, extracts subcircuits consisting of CNOT gates using a circuit sweeping technique, and rewrites each subcircuit with a functionally equivalent and depth-optimal circuit generated by an SAT solver. 
Experimental results demonstrate that, compared with existing post-QCT optimization approaches such as CBIR and Q-Synth, SSR achieves superior performance, improving circuit depth by up to 29.04\% and 16.59\% on average across all benchmark circuits, whereas the depth reduction of CBIR and Q-Synth are, respectively, 12.17\% and 16.63\% at most and 6.45\% and -0.53\% on average on the same benchmarks.
\end{abstract}



\received{20 February 2007}
\received[revised]{12 March 2009}
\received[accepted]{5 June 2009}

\maketitle

\section{Introduction}
Quantum computing has emerged as a transformative technology with the potential to solve classically intractable problems exponentially or polynomially faster than classical approaches~\cite{deutsch1992rapid,shor1994polynomial,harrow2009quantum,grover1996fast}. However, current quantum devices, known as Noisy Intermediate-Scale Quantum (NISQ) devices, suffer from certain physical limitations, such as short qubit lifetime, low gate fidelity, and constrained qubit topology, which pose big challenges for the reliable execution of quantum circuits on these devices.

Quantum circuit transformation (QCT)~\cite{booth2018comparing,venturelli2017temporal,zhang2020depth,li2019tackling,huang2024qubit} involves a series of functionally equivalent transformations on quantum circuits to ensure compliance with the physical constraints of the underlying device, including the native quantum gate set and allowable qubit connectivity. To ensure that each two-qubit gate in a quantum circuit satisfies the qubit connectivity constraints, it may be inevitable for QCT algorithms to insert many SWAP gates into the original circuit. Whether a QCT algorithm is depth-optimal or not, it generally increases the depth of the original circuit, potentially leading to more errors during the execution process. Therefore, it is crucial to offset the circuit-depth overhead introduced during the QCT process through some post-optimization passes. 

Currently, most of the existing work on quantum circuit optimization focuses on quantum circuit synthesis, where the goal is to directly optimize from the logical circuit layer to the physical device~\cite{davis2020towards,xu2023synthesizing,niu2023powerful}. 
In this category, topology-aware synthesis and qubit routing are often formulated as exact or near-exact optimization problems using SAT~\cite{dac19/wille_mapping}, SMT~\cite{Tan2020Optimal}, or ILP~\cite{Almeida+19_permutation} techniques.

Additionally, there are also other works that consider the work of circuit optimisation itself.
Xu et at.~\cite{xu2025optimizing} considered the rewriting rules and unitary resynthesis to approximately optimize quantum circuits. 
Arora et al.~\cite{arora2025local} presented a local optimization algorithm to repeatedly optimize local subcircuits in order to achieve the goal of overall circuit optimization.
However, these approaches do not specifically address the optimization of circuits that have already transformed into a hardware-executable form. The QCT process, which typically involves inserting extra two-qubit gates such as SWAP and CNOT gates to accommodate hardware connectivity constraints, results in circuits that may still be suboptimal in terms of depth, fidelity, or gate efficiency.

In contrast, post-QCT optimization focuses on improving circuits that have already been transformed, considering the specific challenges and constraints imposed by the quantum hardware. A few notable works have targeted this type of optimization. For instance, Wu et al.~\cite{wu2020qgo} proposed QGo, a hierarchical, block-by-block optimization framework to reduce the CNOT gate count. Chen et al.~\cite{chen2022recursive} applied their synthesis method to reduce the number of SWAP gates for the QCT-transformed circuits. 
Q-Synth~\cite{ShaikvdP2024cnotsynthesis} supports layout-restricted CNOT synthesis and includes a peephole optimization module that extracts CNOT slices and optimizes each slice under connectivity constraints. 
Additionally, Itoko et al.~\cite{itoko2019optimization} and Xie et al.~\cite{xie2021mitigating} present commutation-based post-mapping optimization methods, leveraging gate transformations and instruction reordering to improve circuit depth and mitigate crosstalk under connectivity constraints.
These methods aim to refine QCT-transformed circuits to achieve better performance on NISQ devices, focusing on optimizing the transformed circuits to further reduce errors and improve execution fidelity.

Due to the presence of CNOT gates both in the original quantum circuits and added during the QCT process, the executable quantum circuits submitted to the underlying hardware often contain many sub-circuits made up of consecutive CNOT gates\footnote{A SWAP gate can be decomposed to 3 consecutive CNOT gates.}. This creates significant opportunities for optimizing the overall circuit depth.
Focusing on such circuit pattern, we propose a Swapping-Sweeping-and-Rewriting (SSR) optimization framework, which is a significant extension of the SAT-Sweeping technique~\cite{kuehlmann2004dynamic, fujita2015toward, kuehlmann2002robust, mishchenko2005fraigs, mishchenko2006improvements, zhu2006sat} commonly utilized by classical EDA tools, to further reduce the depth of the QCT circuit. We sweep throughout the entire input QCT circuit and extract CNOT subcircuits to be optimized. For any extracted subcircuit, we rewrite it using the corresponding depth-optimal circuit obtained from an SAT solver with a modified CNF encoding scheme. To minimize the number of trials needed for the SAT solver to identify the depth-optimal circuit, we have trained an artificial neural network (ANN) model to predict the optimal depth of any CNOT circuit while considering the qubit connectivity constraints. Moreover, to leverage the generalized communication rules on SWAP and CNOT gates (to be mentioned in Sec.~\ref{sec:swapping_ga}), a genetic algorithm (GA) is proposed to heuristically explore the position exchange opportunities for SWAP and CNOT gates while adhering to qubit connectivity constraints. The GA-based module can offer more depth-optimization possibilities for the subsequent Sweeping and rewriting procedures. The experimental results show that our SSR optimizer can significantly reduce the depth of quantum circuits generated by the QCT process by  29.04\% at most and 16.59\% on average. 

The main contributions of this work are summarized as follows:
\begin{itemize}
\item We formulate depth-oriented post-QCT optimization as a dedicated problem and propose SSR as a unified framework that specifically targets depth reduction of layout-aware quantum circuits, which is complementary to existing QCT and synthesis techniques.
\item We introduce an ANN-assisted depth prediction scheme to accelerate SAT-based CNOT rewriting and evaluating candidate subcircuits. The prediction is used solely to guide the search order and reduce solver invocations, while the SAT procedure itself remains complete and preserves optimality.
\item We propose a gate position constraining (blocked position) mechanism in the SAT formulation to balance local depth optimality of CNOT subcircuits with global circuit-level depth reduction.
\item We design a subcircuit scoring and selection strategy, combined with a GA-based commutation search, which enables SSR to prioritize the most promising subcircuits and iteratively expose new optimization opportunities.
\item Extensive experiments on multiple architectures and benchmark suites demonstrate that SSR achieves consistent and significant depth reductions, with up to 29.04\% improvement and 16.59\% reduction on average.
\end{itemize}

\section{Preliminaries}

\subsection{Quantum circuit}
\paragraph{Qubit and quantum gates}
Qubits are the fundamental units of information in a quantum computer. 
The state of a single qubit is described by a quantum state vector, denoted as $|\psi\rangle$, which can be expressed in the form:
$$
|\psi\rangle = \alpha |0\rangle + \beta |1\rangle,
$$
where $|0\rangle$ and $|1\rangle$ are the computational basis states, analogous to the classical bit values 0 and 1. The coefficients $\alpha$ and $\beta$ are complex numbers, known as probability amplitudes, that satisfy the normalization condition $|\alpha|^2 + |\beta|^2 = 1$.

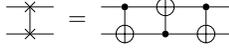
\begin{figure}
    \centering
\scalebox{1.2}{
\begin{tikzpicture}
\begin{yquantgroup}
\registers{
      qubit {} q[2];
   }
\circuit{
    swap (q[0], q[1]);
}
\equals
\circuit{
    cnot q[1] | q[0];
    cnot q[0] | q[1];
    cnot q[1] | q[0];
}
\end{yquantgroup}
\end{tikzpicture}
}
\caption{SWAP gate decomposition}
\label{fig:swap_decp}
\end{figure}

Quantum gates are fundamental operations that manipulate the state of qubits. Current NISQ devices typically only allow for native gate sets consisting of single and two-quit gates. One of the most commonly used two-qubit native gates is the CNOT (Controlled-NOT) gate, denoted by CNOT($q_i, q_j$) in which $q_i$ and $q_j$ are the control and target qubits, respectively. The CNOT gate performs a NOT operation on the target qubit if and only if the control qubit is in the state $|1\rangle$. Another important (but not native) two-qubit gate is the SWAP gate, which exchanges the states of two qubits. The SWAP gate can be decomposed into a sequence of three CNOT gates, as illustrated in Fig.~\ref{fig:swap_decp}

\paragraph{Quantum circuit} A quantum circuit is a graphical model that represents a sequence of quantum operations applied to qubits during the execution of a quantum algorithm. Formally, a quantum circuit can be described as a tuple $(Q, C)$, where $Q = \{q_1, \ldots, q_n\}$ is the set of qubits, and $C = \{g_1, \ldots, g_m\}$ is the ordered sequence of quantum gates applied to these qubits. For simplicity, we use $C$ directly to refer to the quantum circuit.
If a $C$ consists solely of CNOT (and SWAP) gates, the circuit is referred to as a CNOT circuit. 

\subsection{NISQ device}

\begin{figure}
     \centering
     \includegraphics[width=0.4\linewidth]{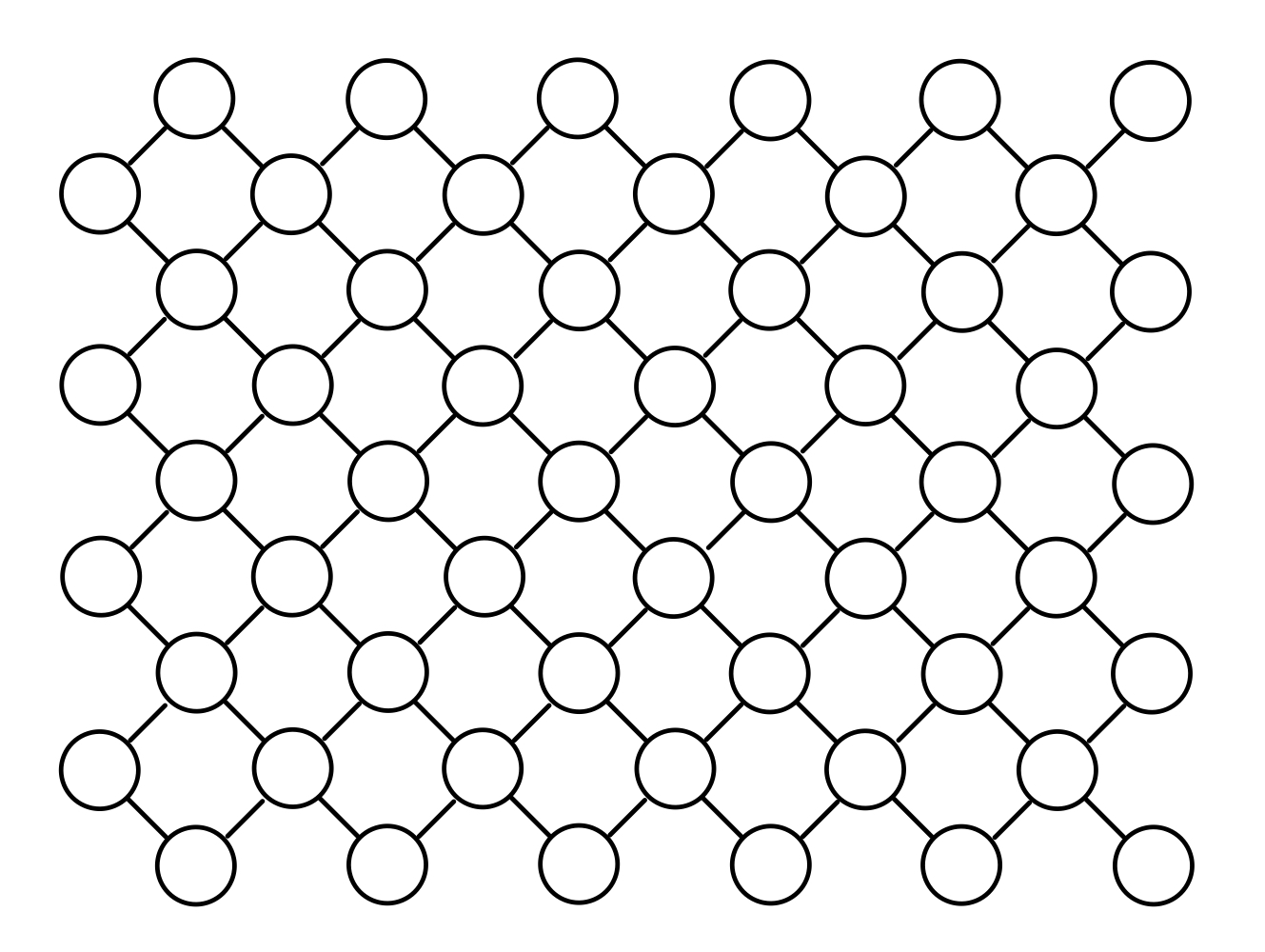}
     \caption{Architecture graph of Google Sycamore}
     \label{fig:sym54}
 \end{figure}

Current NISQ devices are characterized by their limited number of qubits and susceptibility to noise. Additionally, some leading quantum computing technologies such as superconducting circuits impose another significant constraint on physical devices: restricted qubit connectivity. The qubit connectivity of a quantum device can be represented by an undirected graph, termed architecture graph denoted as $AG = (V, E)$, where $V$ denotes the set of physical qubits and $E$ is the set of allowable two-qubit interactions. Fig.~\ref{fig:sym54} presents the architecture graph of the NISQ device known as Google Sycamore~\cite{arute2019quantum}. 

\subsection{Quantum circuit transformation and optimization}

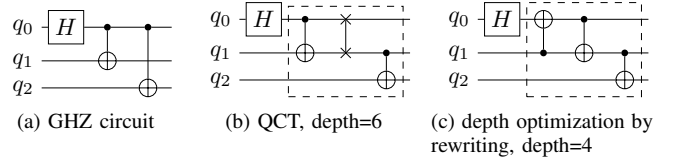
\begin{figure}
    \centering
\scalebox{1.2}{
\subfloat[GHZ circuit]{    
\begin{tikzpicture}
\begin{yquant}
    qubit {$q_{\idx}$}  q[3];
    h q[0];
    cnot q[1] | q[0];
    cnot q[2] | q[0];
    
\end{yquant}
\end{tikzpicture}
}
}
\hfill
\scalebox{1.2}{
\subfloat[QCT, depth=6]{
\begin{tikzpicture}
\begin{yquant*}
    qubit {$q_{\idx}$}  q[3];
    h q[0];
    [name=left1]
    cnot q[1] | q[0];
    [name=left]
    swap (q[0], q[1]);
    [name=righttop]
    cnot q[2] | q[1];
\end{yquant*}
\node[draw, dashed, fit=(left1)(left) (righttop)] {};
\end{tikzpicture}
}
}
\hfill
\scalebox{1.2}{
\subfloat[depth optimization by rewriting, depth=4]{
\begin{tikzpicture}
\begin{yquant}
    qubit {$q_{\idx}$}  q[3];
    h q[0];
    [name=left1]
    cnot q[0] | q[1];
    cnot q[1] | q[0];
    [name=righttop]
    cnot q[2] | q[1];
\end{yquant}
\node[draw, dashed, fit=(left1)(left) (righttop)] {};
\end{tikzpicture}
}
}
\caption{Quantum circuit transformation and depth-optimized rewriting. The AG is linear nearest neighbor and $q_i$ is mapped to physical qubit $v_i$.
}
    \label{fig:qct_rw}
\end{figure}

The restricted connectivity constraint states that each two-qubit gate can only operate on a pair of directly coupled qubits in the target device, which corresponds to an edge in the specific AG. When a two-qubit gate in a quantum circuit does not comply with this connectivity constraint, QCT is required to `move' its two virtual operand qubits to a pair of directly coupled physical qubits on the device by inserting SWAP gates. We will refer to the circuit after QCT as QCT circuit hereinafter.
Fig.~\ref{fig:qct_rw} illustrates an example of QCT, where a GHZ circuit (a),
i.e., a three-qubit Greenberger–Horne–Zeilinger state preparation circuit consisting of one Hadamard gate followed by two CNOT gates,
is adapted to fit a linear qubit topology using one SWAP gate (b). From this simple example, it is evident that QCT circuits, particularly those containing a significant number of two-qubit gates originally, will exhibit multiple sub-circuits solely composed of CNOT gates (SWAP gates will be decomposed to CNOT gates as shown in Fig.~\ref{fig:swap_decp}). If we rewrite these CNOT subcircuits, that is, replacing them with equivalent circuits that have smaller depths, it is very likely that the overall depth of the entire circuit will be reduced as well (cf. Fig.~\ref{fig:qct_rw}(c)). 
It should be noted that the connectivity constraint should also be considered when performing any post-QCT optimization.

\subsection{CNOT circuits and CNF formulations}
\label{sec:cnf_form}

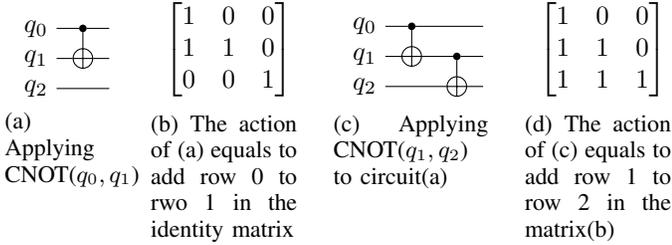
\begin{figure}
\centering
\scalebox{1.1}{
\subfloat[Applying CNOT($q_0,q_1$)]{
\begin{tikzpicture}
\begin{yquant}
    qubit {$q_{\idx}$} q[3];
    cnot q[1] | q[0]; 
\end{yquant}
\end{tikzpicture}
}}
\hfill
\scalebox{1.1}{
\subfloat[The action of (a) is equivalent to adding row 0 to row 1 in the identity matrix ]{
$\begin{bmatrix}
1 & 0 & 0\\
1 & 1 & 0 \\
0 & 0 & 1
\end{bmatrix}$
}}
\hfill
\scalebox{1.1}{
\subfloat[Applying CNOT($q_1,q_2$) to circuit in (a)]{
\begin{tikzpicture}
\begin{yquant}
    qubit {$q_{\idx}$} q[3];
    cnot q[1] | q[0];
    cnot q[2] | q[1];
\end{yquant}
\end{tikzpicture}
}}
\hfill
\scalebox{1.1}{
\subfloat[The action of (c) is equivalent to adding row 1 to row 2 in the matrix in (b)]{
$\begin{bmatrix}
1 & 0 & 0 \\
1 & 1 & 0 \\
1 & 1 & 1
\end{bmatrix}$
}}
\hfill
\scalebox{1.1}{
\subfloat[Applying SWAP($q_0,q_1)$ to circuit in (c)]{
\begin{tikzpicture}
\begin{yquant}
    qubit {$q_{\idx}$} q[3];
    cnot q[1] | q[0];
    cnot q[2] | q[1];
    swap (q[0], q[1]);
\end{yquant}
\end{tikzpicture}
}
}
\hfill
\scalebox{1.1}{
\subfloat[The action of (e) is equivalent to exchange row 0 and row 1 in the matrix in (d)]{
$\begin{bmatrix}
1 & 1 & 0 \\
1 & 0 & 0 \\
1 & 1 & 1
\end{bmatrix}$
}}
\caption{Matrix representation of CNOT circuit}
\label{fig:cnot_matrix}
\end{figure}

In boolean algebra, a CNOT circuit with $n$ qubits can be depicted as an $n \times n$ invertible matrix $M$ over the field $\mathbb{F}_2$ (refer to Fig.~\ref{fig:cnot_matrix} for a straightforward example). This matrix-form representation allows for efficient analysis and manipulation of the structure of the circuit, which is essential for optimizing quantum circuits concerning gate count and depth~\cite{jiang2020optimal}. Moreover, the action of a CNOT gate, say CNOT($q_i,q_j$), can be described as an invertible linear transformation over the finite field $\mathbb{F}_2$~\cite{markov2008optimal}. This transformation corresponds to a row operation where the $i$-th row is added to the $j$-th row in $M$.

One significant issue to be tackled in this work is: given any CNOT circuit, is it possible to derive a functionally equivalent circuit with reduced depth? If so, what is the optimal solution constitution? As shown in Ref.~\cite{chen2022optimizing}, this challenge is framed as the Boolean satisfiability (SAT) problem. Meanwhile, a conjunctive normal form (CNF) encoding is presented, translating the task of finding a CNOT circuit with a designated depth into determining satisfiable assignments for all Boolean variables within the CNF. This translation allows the use of existing SAT tools. In our study, we adopt their method for creating the CNF formulations with several major modifications being imposed (cf. Sec.~\ref{sec:disable}). Initially, we convert the target CNOT circuit to the corresponding invertible matrix $M$. The Boolean variable $m_{i,k}^d$ signifies whether the matrix entry at $M_{i,k}$ is 0 or 1 at depth $d$, while $g_{c \rightarrow t}^d$ indicates the presence of the gate CNOT($q_c$, $q_t$) at depth $d$.  The types of clauses in the CNF formulation that serve to constrain the problem are summarized as follows. Readers can refer to Sec. 2 in~\cite{chen2022optimizing} for more detailed descriptions.

\begin{enumerate}
    
    \item  Each qubit can be involved in at most one CNOT at each depth. This prevents the occurrence of overlapping CNOT gates at the same timestep.
   
    \item  If CNOT$(q_i,q_k)$ is applied at depth $d$, then at depth $d+1$, every entry of $M$ in row $k$ must be XOR-ed with the corresponding entry in row $i$ at depth $d$. This captures the impact of the CNOT operation within matrix representation.
  
    \item  If an entry of $M$ in row $i$ changes between depths $d$ and $d+1$,  then exactly one CNOT with target qubit $i$ must exist at depth $d$. This ensures that each entry change in $M$ is explicitly linked to a CNOT gate.
     
\end{enumerate}

In addition to the above constraints, hardware connectivity constraints are explicitly incorporated into our SAT formulation. 
\begin{enumerate}[start=4]
    \item Specifically, we assume a given architecture graph $AG = (V, E)$ that characterizes the native two-qubit connectivity of the target NISQ device. A CNOT gate CNOT$(q_c, q_t)$ is allowed at depth $d$ if and only if $(c,t)$ corresponds to an edge in the architecture graph, i.e., $(c,t) \in E$.
\end{enumerate}
 Accordingly, Boolean variables $g_{c \rightarrow t}^d$ are defined only for hardware-adjacent qubit pairs, while variables corresponding to non-adjacent pairs are omitted from the formulation.
 This effectively prunes infeasible two-qubit interactions from the search space and guarantees that any satisfying assignment of the CNF corresponds to a CNOT circuit compliant with the underlying hardware connectivity. By embedding the architectural constraints directly into the variable domain rather than enforcing them through additional clauses, the size of the CNF instances is reduced and the SAT solver is guided toward hardware-executable solutions. 

Consider, for instance, the target circuit is shown in Fig.~\ref{fig:cnot_matrix}(e), the architecture graph is linear nearest neighbor, meaning connections exist between $(q_0, q_1)$ and $(q_1, q_2)$, and the depth of the circuit to be constructed is 5 (the SWAP gate is decomposed into 3 CNOT gates). Hence, the initial matrix variables are:
$$
\left[\begin{array}{lll}
m_{0,0}^0 & m_{0,1}^0 & m_{0,2}^0 \\
m_{1,0}^0 & m_{1,1}^0 & m_{1,2}^0 \\
m_{2,0}^0 & m_{2,1}^0 & m_{2,2}^0
\end{array}\right] = { }\left[\begin{array}{ccc}
1 & 0 & 0 \\
0 & 1 & 0 \\
0 & 0 & 1
\end{array}\right].
$$
Fig.~\ref{fig:cnot_matrix} illustrates the process of transforming an initial matrix into the target matrix. Specifically, applying a CNOT$(q_c, q_t)$ updates the matrix by XOR-ing row $c$ into row $t$, while applying a SWAP$(q_c, q_t)$ updates the matrix by exchanging rows $c$ and $t$.

Accordingly, Fig.~\ref{fig:cnot_matrix}(f) presents the {target} matrix-form representation.
As a result, the target matrix variables at the final depth are given by
$$
\left[\begin{array}{lll}
m_{0,0}^D & m_{0,1}^D & m_{0,2}^D \\
m_{1,0}^D & m_{1,1}^D & m_{1,2}^D \\
m_{2,0}^D & m_{2,1}^D & m_{2,2}^D
\end{array}\right] = { }\left[\begin{array}{ccc}
1 & 1 & 0 \\
1 & 0 & 0 \\
1 & 1 & 1
\end{array}\right].
$$

To identify a depth-optimal circuit, we initially set the target depth to $D = 1$.
At the first depth, 
the variables $g^0_{0 \to 2}$ and $g^0_{2 \to 0}$ do not exist because of constraint 4.
Concurrently, constraint 1 indicates that $g^0_{0\to 1} + g^0_{1 \to 0} + g^0_{1 \to 2} + g^0_{2\to 1} = 1$. In this case, the overlapped CNOT gates at the same depth, e.g., $g^0_{0\to 1} + g^0_{1 \to 2} > 1$, are prohibited.

Comparing the matrices at depth 0 (initial matrix) and depth 1 (the target matrix since we set $D = 1$), it can be observed that row 1 has changed. 
Therefore, according to constraint 3, we enforce $g^0_{0 \to 1} + g^0_{2 \to 1} = 1$.
Now assume $g^0_{0\to 1} = 1$. 
Then, according to constraint 2,
the matrix variables at depth 1 are updated as follows:
$$
\begin{aligned}
&\left[\begin{array}{lll}
m_{0,0}^0 & m_{0,1}^0 & m_{0,2}^0 \\
m_{1,0}^0 & m_{1,1}^0 & m_{1,2}^0 \\
m_{2,0}^0 & m_{2,1}^0 & m_{2,2}^0
\end{array}\right] \\
\xrightarrow[g_{0 \rightarrow 1}^0]{} &\left[\begin{array}{lll}
m_{0,0}^0 & m_{0,1}^0 & m_{0,2}^0 \\
m_{0,0}^0 \oplus m_{1,0}^0 & m_{0,1}^0 \oplus m_{1,1}^0 & m_{0,2}^0 \oplus m_{1,2}^0 \\
m_{2,0}^0 & m_{2,1}^0 & m_{2,2}^0 
\end{array}\right] \\
=& \left[\begin{array}{lll}
m_{0,0}^1 & m_{0,1}^1 & m_{0,2}^1 \\
m_{1,0}^1 & m_{1,1}^1 & m_{1,2}^1 \\
m_{2,0}^1 & m_{2,1}^1 & m_{2,2}^1
\end{array}\right] = 
\left[\begin{array}{ccc}
1 & 0 & 0 \\
1 & 1 & 0 \\
0 & 0 & 1
\end{array}\right],
\end{aligned}
$$
which contradicts the target matrix in Fig.~\ref{fig:cnot_matrix} (f). The same contradiction occurs for $g^0_{2\to 1}$.
As a result, no satisfying assignment exists for target depth $D = 1$.

If the SAT solver cannot find a feasible assignment for a given depth limit, the target depth $D$ is gradually increased until a satisfiable solution is found. This commonly adopted strategy employs the SAT solver in a trial-and-error fashion. Although it ensures optimality, it can lead to significant computational cost when the optimal depth is large. In addition, the original formulation optimizes subcircuits independently, which does not necessarily reduce the total circuit depth because of interactions with neighboring gates ({cf. Fig.~9 in Ref.~\cite{chen2022recursive}}).

To address these limitations, we introduce (i) an ANN-assisted strategy to guide the invocation of the SAT solver towards promising depth candidates, and (ii) a gate position constraining mechanism that incorporates global circuit context into the CNF encoding. The former significantly reduces the number of unsuccessful SAT invocations, while the latter {effectively} prevents locally optimal yet globally ineffective rewritings. Together, these two modifications enable a more efficient and depth-aware SAT-based circuit rewriting process.

\section{Methodology}

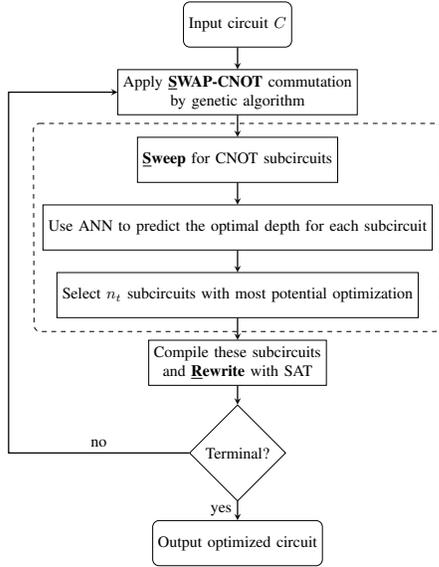
\begin{figure}
\centering
\scalebox{0.8}{
\begin{tikzpicture}[node distance=1.5cm]
\node (start) [startstop] {Input circuit $C$};
\node (genetic) [process, below of=start,align=center] {Apply \textbf{\underline{S}WAP-CNOT} commutation \\ by genetic algorithm};
\node (sweep) [process, below of=genetic] {\textbf{\underline{S}weep} for CNOT subcircuits};
\node (ANN) [process, below of=sweep] {Use ANN to predict the optimal depth for each subcircuit};
\node (select) [process, below of=ANN] {Select $n_t$ subcircuits with most potential optimization};
\node (SAT) [process, below of=select,yshift=-0.5cm,align=center] {Compile these subcircuits \\ and \textbf{\underline{R}ewrite} with SAT \\
guided by ANN
};
\node (terminal) [decision, below of=SAT,yshift=-0.5cm,align=center] {Terminal?};

\node (end) [startstop, below of=terminal,yshift=-0.5cm] {Output optimized circuit};

\draw [arrow] (start) -- (genetic);
\draw [arrow] (genetic) -- (sweep);
\draw [arrow] (sweep) -- (ANN);
\draw [arrow] (ANN) -- (select);
\draw [arrow] (select) -- (SAT);
\draw [arrow] (SAT) -- (terminal);
\draw [arrow] (terminal) -- node[anchor=east]{yes}(end);
\draw [arrow] (terminal.west) --node[above]{no} ++(-4,  0) |- (genetic.west);
\begin{scope}[]
    \draw[dashed, rounded corners] 
        ($(sweep.north west) + (-2.3, 0.3)$) -- 
        ($(sweep.north east) + (2.3, 0.3)$) -- 
        ($(select.south east) + (0.5, -0.3)$) -- 
        ($(select.south west) + (-0.5, -0.3)$) -- cycle;
\end{scope}
\end{tikzpicture}
}
\caption{The flowchart of SSR, the dotted box represents the whole procedure of subcircuits sweeping. }
\label{fig:flowchart}
\end{figure}

In this section, we formally introduce the \underline{S}wapping-\underline{S}weeping-and-\underline{R}ewriting (SSR)
framework to optimize QCT circuits.
We assume that any input quantum circuit to SSR is a hardware-compliant circuit compiled by the QCT process.
The flowchart of our SSR algorithm is shown in Fig.~\ref{fig:flowchart}, where we iteratively perform the following steps, including SWAP commutation, subcircuit sweeping, and SAT-based rewriting,  until the overall depth of the optimized circuit cannot be further reduced.  
Meanwhile, Alg.~\ref{alg:ssr} provides a detailed description of the SSR procedure.

\begin{algorithm}
\caption{SSR: Depth-Oriented Post-QCT Optimization Framework}
\label{alg:ssr}
\begin{algorithmic}[1]
\REQUIRE Mapped quantum circuit $\mathcal{C}_{\text{init}}$, 
maximum iterations $T$, 
\ENSURE Optimized circuit $\mathcal{C}^{*}$
\STATE $\mathcal{C} \leftarrow \mathcal{C}_{\text{init}}$
\FOR{$t = 1$ to $T$}
    \STATE $s$ = GABasedCommutation($\mathcal{C}$, $n_{\text{species}}$, $\alpha$, $\alpha_\mu$, $T_{\text{max}}$, $T_{\text{idle}}$)  // Apply SWAP-CNOT commutation by Genetic Algorithm. 
    \STATE Obtain the best candidate circuit  $\mathcal{C}_{\text{GA}}$ according to the solution $s$
    \STATE $\mathcal{C} \leftarrow \mathcal{C}_{\text{GA}}$
    \STATE Extract CNOT subcircuits $\{\mathcal{C}_1, \mathcal{C}_2, \ldots, \mathcal{C}_m\}$ from $C$ by subcircuit sweeping
    \FORALL{subcircuits $\mathcal{C}_i$}
        \STATE Predict depth $d^i_{\text{pred}}$ of subcircuit $\mathcal{C}_i$ using ANN
        \STATE Compute optimization score ${score}(\mathcal{C}_i)$
    \ENDFOR

    \STATE Select top-$n_t$ subcircuits with highest scores
    \FORALL{selected subcircuits $\mathcal{C}_i$}
        \STATE $\mathcal{C}_{i_{\text{opt}}} \leftarrow $ SATBasedRewriting($\mathcal{C}_{\text{ori}}$, $\mathcal{C}$, $d^i_\text{pred}$)  // Apply SAT-based rewriting with hardware and position constraints. 
        \IF{SAT returns a valid rewrite}
            \STATE Update $\mathcal{C}$ by replacing $\mathcal{C}_i$
        \ENDIF
    \ENDFOR
\ENDFOR
\STATE \textbf{return} $\mathcal{C}^{*} \leftarrow \mathcal{C}$
\end{algorithmic}
\end{algorithm}

\subsection{Generalized commutations with genetic algorithm}
\label{sec:swapping_ga}

\paragraph{Generalized Commutation Rules.}

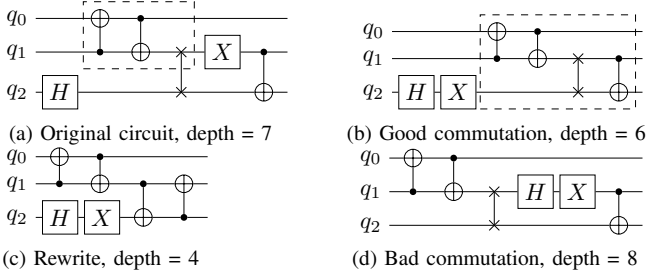
\begin{figure}
    \centering

\scalebox{0.9}{
\subfloat[Original circuit. The SWAP gate should be regarded as three CNOT. Therefore, the depth = 7.]{
\begin{tikzpicture}
\begin{yquant*}
    qubit {$\reg_{\idx}$} q[3];
    h q[2];
    null q[0];
    [name=left1]
    cnot q[0] | q[1];
    [name=left2]
    cnot q[1] | q[0];
    [name=left3]
    swap (q[1], q[2]);
    x q[1];
    [name=right1]
    cnot q[2] | q[1];
    [name=right2]
\end{yquant*}
\node[draw, dashed, fit=(left1) (left2)(left3)]{};
\end{tikzpicture}
}}
\hfill
\scalebox{0.9}{
\subfloat[Good commutation, i.e., exchanging the X and the SWAP gate in the circuit in Fig~\ref{fig:circ_extract_swap}(a). Then the circuit depth is reduced to 6.]{
\begin{tikzpicture}
\begin{yquant*}
    qubit {$\reg_{\idx}$} q[3];
    h q[2];
    x q[2];
    null q[0];
    null q[0];
    [name=left1]
    cnot q[0] | q[1];
    cnot q[1] | q[0];
    [name=left3]
    swap (q[1], q[2]);
    [name=left2]
    cnot q[2] | q[1];
\end{yquant*}
\node[draw, dashed, fit=(left1) (left2)(left3)]{};
\end{tikzpicture}
}
}
\hfill
\scalebox{0.9}{
\subfloat[Rewrite the CNOT subcircuit enclosed in the dashed line in Fig~\ref{fig:circ_extract_swap}(a). Then the circuit depth is reduced to  4.]{
\begin{tikzpicture}
\begin{yquant*}
    qubit {$\reg_{\idx}$} q[3];
    h q[2];
    x q[2];  
    cnot q[0] | q[1];
    cnot q[1] | q[0];
    cnot q[2] | q[1];
    cnot q[1] | q[2];
\end{yquant*}
\end{tikzpicture}
}
}
\hfill
\scalebox{0.9}{
\subfloat[Bad commutation, i.e., exchanging the H and the SWAP gate  in the circuit in Fig~\ref{fig:circ_extract_swap}(a). Then the circuit depth is increased to  8.]{
\begin{tikzpicture}
\begin{yquant*}
    qubit {$\reg_{\idx}$} q[3];
    cnot q[0] | q[1];
    cnot q[1] | q[0];
    swap (q[1], q[2]);
    h q[1];
    x q[1];
    cnot q[2] | q[1];
\end{yquant*}
\end{tikzpicture}
}
}
\caption{Examples for applying SWAP commutation rules and CNOT circuit sweeping.}
\label{fig:circ_extract_swap}
\end{figure}

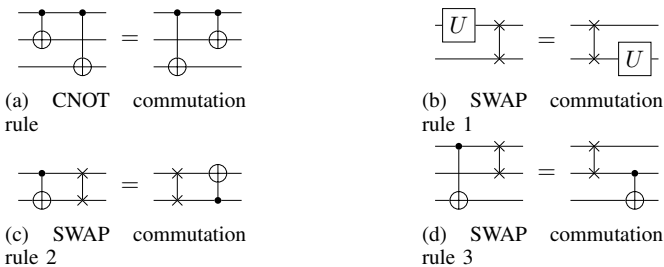
\begin{figure}
\centering
\scalebox{0.9}{
\subfloat[CNOT commutation rule]{
\begin{tikzpicture}
\begin{yquantgroup}
\registers{
      qubit {} q[3];
   }
\circuit{
    cnot q[1] | q[0];
    cnot q[2] | q[0];
}
\equals
\circuit{
    cnot q[2] | q[0];
    cnot q[1] | q[0];
}
\end{yquantgroup}
\end{tikzpicture}
} }
\hfill
\scalebox{0.9}{
\subfloat[SWAP commutation rule 1]{
\begin{tikzpicture}
\begin{yquantgroup}
\registers{
      qubit {} q[2];
   }
\circuit{
    box {$U$} q[0];
    swap (q[0],q[1]);
}
\equals
\circuit{
    swap (q[0],q[1]);
    box {$U$} q[1];
}
\end{yquantgroup}
\end{tikzpicture}
}}
\hfill
\scalebox{0.9}{
\subfloat[SWAP commutation rule 2]{
\begin{tikzpicture}
\begin{yquantgroup}
\registers{
      qubit {} q[2];
   }
\circuit{
    cnot q[1] | q[0];
    swap (q[0],q[1]);
}
\equals
\circuit{
    swap (q[0],q[1]);
    cnot q[0] | q[1];
}
\end{yquantgroup}
\end{tikzpicture}
}}
\hfill
\scalebox{0.9}{
\subfloat[SWAP commutation rule 3]{
\begin{tikzpicture}
\begin{yquantgroup}
\registers{
      qubit {} q[3];
   }
\circuit{
    cnot q[2] | q[0];
    swap (q[0],q[1]);
}
\equals
\circuit{
    swap (q[0],q[1]);
    cnot q[2] | q[1];
}
\end{yquantgroup}
\end{tikzpicture}
}}
\caption{Generalized commutation rules of SWAP and CNOT \cite{xie2021mitigating}.}
\label{fig:comm_rule}
\end{figure}

For QCT circuits, there may exist multiple SWAP and CNOT gates on which a series of generalized commutation rules (cf. Sec. III-A in ~\cite{xie2021mitigating} ) can be applied.
If we rearrange the quantum gates in a quantum circuit based on their commutativity, the depth of the entire circuit as well as the configuration of the sub-circuits to be extracted will change accordingly, which may provide more opportunities for minimizing the circuit depth.
For example, in Fig.~\ref{fig:circ_extract_swap}(a), the exchange between the X and the SWAP gate directly leads to a depth reduction from 7 to 6 (cf. Fig.~\ref{fig:circ_extract_swap}(b)), and this depth can be further reduced to 4 after rewriting the subcircuit on the right (cf. Fig.~\ref{fig:circ_extract_swap}(c)) via the technique to be introduced in Sec.~\ref{sec:disable}. 
Since the connectivity constraints of the underlying devices require that two-qubit gates can only act on those directly coupled physical qubits, any exchange of two quantum gates that violates the connectivity constraint is prohibited.
In Fig.~\ref{fig:circ_extract_swap}(a), the  CNOT gate on the left side of SWAP is not allowed to interchange with the SWAP if the AG is a path graph, i.e., $q_0,q_1$, and $q_2$ being arranged in a linear topology. 
It is worth noting that we only consider the generalized commutation rules based on SWAP and CNOT gates in Fig.~\ref{fig:comm_rule}, since the proposed framework is tailored for QCT circuits.

\paragraph{Genetic Algorithm}

For an input circuit, there may be multiple ways to apply the commutation rules and not all of them are beneficial. For example, if we exchange the H gate and the SWAP gate in Fig.~\ref{fig:circ_extract_swap}(a), the depth of the resulting circuit will become 8 (cf. Fig.~\ref{fig:circ_extract_swap}(d)).
The problem of identifying the commutation sequence based on rules depicted in Fig.~\ref{fig:comm_rule} can be considered as a typical combinatorial optimization problem, which can be efficiently solved via the genetic algorithm (GA)~\cite{lambora2019genetic}. GA is an optimization technique inspired by the principles of natural evolution and genetics. GA has also been widely used applied in several other domains related to various aspects of quantum computing, such as quantum state preparation~\cite{creevey2023gasp} and quantum simulation~\cite{las2016genetic}. The key components of GA include \emph{selection}, \emph{crossover}, and \emph{mutation}, through the combination of which the population can be driven toward increasingly optimal solutions.

Given an input quantum circuit $\mathcal{C}$ consisting of $N$ gates, we assign each gate a unique index according to its initial topological order, and denote the $i$-th gate by $g_i$:
$$
\mathcal{C}_0 = \{g_1, g_2, ..., g_{N}\}.
$$
In the proposed SSR framework, the solution space comprises all circuits that can be reached from $\mathcal{C}_0$ through a sequence of valid gate commutations. Instead of directly encoding a final gate permutation, we represent each chromosome as a \textit{sequence of commutation operations} denoted as
$$
s=\{(g_{i_1}, g_{j_1}),(g_{i_2}, g_{j_2}), \cdots,(g_{i_L}, g_{j_L})\},
$$
where each pair $(g_{i_k}, g_{i_j})$ represents a commutation operation between the two gates.
Note that such a commutation is considered legal only if $g_{i_k}$ and $g_{j_k}$ comply with one rule in Fig.~\ref{fig:comm_rule} after all its previous gate pairs in $s$ have been commuted.

Obviously, applying the sequence $s$ to the initial circuit $\mathcal{C}_0$ produces a reordered circuit $\mathcal{C}(s)$. In practice, the qubits on which 2-qubit gates act can vary as commutations are applied, which may yield 2-qubit gates violating the hardware connectivity constraints. In our framework, commutation operations are restricted by a predefined set of four commutation rules shown in Fig.~\ref{fig:comm_rule}. Hardware connectivity constraints arise only in the fourth commutation rule, which involves the commutation between a SWAP gate and a CNOT gate that share exactly one common qubit. Suppose we apply this rule to SWAP$(q_i, q_j)$ and CNOT$(q_j, q_k)$. In this case, the commutation is permitted only if qubits $q_i$ and $q_k$ are directly connected in the underlying architecture graph. Otherwise, the resulting CNOT operation would violate hardware connectivity and is therefore disallowed.

To prevent the generation of illegal solutions, such invalid commutation pairs are filtered out during chromosome construction and mutation. Consequently, all chromosomes generated by the GA inherently satisfy hardware connectivity constraints, and no additional repair or rejection step is required. In practice, this design prevents the occurrence of invalid candidates, ensuring that the GA search is confined to the space of hardware-compliant circuits.

\begin{figure*}
\centering
\subfloat[The architecture graph used in this example.]{
\includegraphics[width=0.2\textwidth]{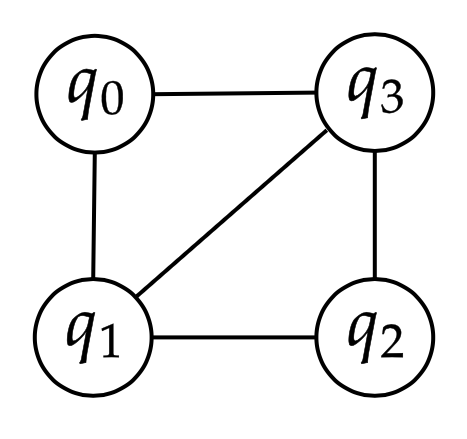}
}
\hfill
\scalebox{1.1}{
\subfloat[The input circuit before GA-based commutation]{
\begin{tikzpicture}
\begin{yquant*}
    qubit { $\reg_{\idx}$} q[4];
    nobit depth;
    h q[2];
    null q[0];
    cnot q[0] | q[1];
    cnot q[1] | q[0];
    cnot q[2] | q[1];
    swap (q[2],q[3]);
    x q[2];
    swap (q[1], q[2]);
    cnot q[3] | q[2];
    cnot q[2] | q[1];

    [draw=none] box {$g_0$} depth;
    [draw=none] box {$g_1$} depth;
    [draw=none] box {$g_2$} depth;
    [draw=none] box {$g_3$} depth;
    [draw=none] box {$g_4$} depth;
    [draw=none] box {$g_5$} depth;
    [draw=none] box {$g_6$} depth;
    [draw=none] box {$g_7$} depth;
    [draw=none] box {$g_8$} depth;   
\end{yquant*}
\end{tikzpicture}
}}

\hfill

\scalebox{1.1}{
\subfloat[Solution $s_1$:$\{(g_3, g_4),(g_4, g_0)\}$, generated during the initialization phase by consecutively applying commutation rules to the input circuit, first exchanging gates $g_3$, $g_4$ and then exchanging gates $g_4$, $g_0$.]{
\begin{tikzpicture}
\begin{yquant*}
    qubit { $\reg_{\idx}$} q[4];
    nobit depth;
    cnot q[0] | q[1];
    null q[2];
    swap (q[2], q[3]);
    null q[0];
    cnot q[1] | q[0];
    null q[3];
    h q[3];    
    cnot q[3] | q[1];
    x q[2];
    swap (q[1],q[2]);
    cnot q[3] | q[2];
    cnot q[2] | q[1];

    [draw=none] box {$g_1$} depth;
    [draw=none] box {$g_4$} depth;
    [draw=none] box {$g_2$} depth;
    [draw=none] box {$g_0$} depth;
    [draw=none] box {$g_3$} depth;
    [draw=none] box {$g_5$} depth;
    [draw=none] box {$g_6$} depth;
    [draw=none] box {$g_7$} depth;
    [draw=none] box {$g_8$} depth;   
\end{yquant*}
\end{tikzpicture}
}}
\hfill
\scalebox{1.1}{
\subfloat[Solution $s_2:\{(g_4, g_5),(g_6,g_7)\}$ , generated during the initialization phase by consecutively applying commutation rules to the input circuit, first exchanging gates $g_4$, $g_5$ and then exchanging gates $g_6$, $g_7$.]{
\begin{tikzpicture}
\begin{yquant*}
    qubit { $\reg_{\idx}$} q[4];
    nobit depth;
    h q[2];
    null q[0];
    cnot q[0] | q[1];
    cnot q[1] | q[0];
    cnot q[2] | q[1];
    null q[3];
    null q[3];
    null q[3];
    null q[3];
    x q[3];
    swap (q[2],q[3]);
    cnot q[3] | q[1];
    swap (q[1], q[2]);
    cnot q[2] | q[1];

    [draw=none] box {$g_0$} depth;
    [draw=none] box {$g_1$} depth;
    [draw=none] box {$g_2$} depth;
    [draw=none] box {$g_3$} depth;
    [draw=none] box {$g_5$} depth;
    [draw=none] box {$g_4$} depth;
    [draw=none] box {$g_7$} depth;
    [draw=none] box {$g_6$} depth;
    [draw=none] box {$g_8$} depth;   
\end{yquant*}
\end{tikzpicture}
}}
\hfill
\scalebox{1.1}{
\subfloat[Solution $s_3:\{(g_6,g_7)\}$, generated during the initialization phase by consecutively applying commutation rules to the input circuit, exchanging gates $g_6$, $g_7$.]{
\begin{tikzpicture}
\begin{yquant*}
    qubit { $\reg_{\idx}$} q[4];
    nobit depth;
    h q[2];
    null q[0];
    cnot q[0] | q[1];
    cnot q[1] | q[0];
    cnot q[2] | q[1];
    swap (q[2],q[3]);
    x q[2];
    null q[3];
    cnot q[3] | q[1];
    swap (q[1], q[2]);
    cnot q[2] | q[1];

    [draw=none] box {$g_0$} depth;
    [draw=none] box {$g_1$} depth;
    [draw=none] box {$g_2$} depth;
    [draw=none] box {$g_3$} depth;
    [draw=none] box {$g_4$} depth;
    [draw=none] box {$g_5$} depth;
    [draw=none] box {$g_7$} depth;
    [draw=none] box {$g_6$} depth;
    [draw=none] box {$g_8$} depth;   
\end{yquant*}
\end{tikzpicture}
}}
\hfill
\scalebox{1.1}{
\subfloat[Solution $s_4$: $\{(g_4,g_5),(g_6,g_7),(g_4,g_7),(g_6,g_8)\}$, generated via mutation from $s_2$ through the introduction of additional valid commutation operations involving gate pairs $(g_4, g_7)$ and $(g_6, g_8)$.]{
\begin{tikzpicture}
\begin{yquant*}
    qubit { $\reg_{\idx}$} q[4];
    nobit depth;
    h q[2];
    null q[0];
    cnot q[0] | q[1];
    cnot q[1] | q[0];
    cnot q[2] | q[1];
    null q[3];
    null q[3];
    null q[3];
    null q[3];
    x q[3];
    null q[2];
    cnot q[2] | q[1];
    swap (q[2],q[3]);
    cnot q[1] | q[2];
    swap (q[1], q[2]);

    [draw=none] box {$g_0$} depth;
    [draw=none] box {$g_1$} depth;
    [draw=none] box {$g_2$} depth;
    [draw=none] box {$g_3$} depth;
    [draw=none] box {$g_5$} depth;
    [draw=none] box {$g_7$} depth;
    [draw=none] box {$g_4$} depth;
    [draw=none] box {$g_8$} depth;
    [draw=none] box {$g_6$} depth;   
\end{yquant*}
\end{tikzpicture}
}}
\hfill
\scalebox{1.1}{
\subfloat[Solution $s_5$: $\{(g_6,g_7),(g_6,g_8)\}$, generated via mutation from $s_3$ through the introduction of additional valid commutation operations involving gate pair $(g_6, g_7)$.]{
\begin{tikzpicture}
\begin{yquant*}
    qubit { $\reg_{\idx}$} q[4];
    nobit depth;
    h q[2];
    null q[0];
    cnot q[0] | q[1];
    cnot q[1] | q[0];
    cnot q[2] | q[1];
    swap (q[2],q[3]);
    x q[2];
    null q[3];
    cnot q[3] | q[1];
    cnot q[1] | q[2];
    swap (q[1], q[2]);

    [draw=none] box {$g_0$} depth;
    [draw=none] box {$g_1$} depth;
    [draw=none] box {$g_2$} depth;
    [draw=none] box {$g_3$} depth;
    [draw=none] box {$g_4$} depth;
    [draw=none] box {$g_5$} depth;
    [draw=none] box {$g_7$} depth;
    [draw=none] box {$g_8$} depth;
    [draw=none] box {$g_6$} depth;   
\end{yquant*}
\end{tikzpicture}
}}
\hfill
\scalebox{1.1}{
\subfloat[Solution $s_6$: $\{(g_3,g_4),(g_4,g_0),(g_6,g_7)\}$, obtained through a crossover operation where solution $s_1$ is used as the base and a valid commutation operation from $s_2$ is incorporated.]{
\begin{tikzpicture}
\begin{yquant*}
    qubit { $\reg_{\idx}$} q[4];
    nobit depth;
    cnot q[0] | q[1];
    cnot q[1] | q[0];
    null q[2];
    null q[2];
    swap (q[2], q[3]);
    h q[3];
    cnot q[3] | q[1];
    x q[2];
    null q[1];
    cnot q[3] | q[1];
    swap (q[1], q[2]);
    cnot q[2] | q[1];

    [draw=none] box {$g_1$} depth;
    [draw=none] box {$g_2$} depth;
    [draw=none] box {$g_4$} depth;
    [draw=none] box {$g_0$} depth;
    [draw=none] box {$g_3$} depth;
    [draw=none] box {$g_5$} depth;
    [draw=none] box {$g_7$} depth;
    [draw=none] box {$g_6$} depth;
    [draw=none] box {$g_8$} depth;   
\end{yquant*}
\end{tikzpicture}
}
}
\caption{An illustrative example of the GA-based commutation process in SSR.}
\label{fig:ga_eg}
\end{figure*}

Alg.~\ref{alg:ga} presents the full GA procedure, with its primary functionalities being summarized as follows:
\begin{itemize}
    \item \textbf{Initialization:} Given an input circuit, commutation rules are applied randomly to generate new candidate solutions. This process is repeated $3 \times n_{\text{species}}$ times to create the initial population (Line 1 in Alg.~\ref{alg:ga}). 
    \begin{example}
    Suppose the input circuit shown in Fig.~\ref{fig:ga_eg}(b), and the architecture graph shown in Fig.~\ref{fig:ga_eg}(a) are given, and let the number of species be set to $n_{\text{species}} = 3$. We then randomly generate $3 \times n_{\text{species}} = 9$ candidate solutions. We show 3 candidate solutions denoted as $s_1$, $s_2$, and $s_3$, as illustrated in Fig.~\ref{fig:ga_eg}(c)–(e). For the candidate solution $s_1$, we first apply SWAP commutation rule~4 to exchange the SWAP gate $g_4$ with the CNOT gate $g_3$. Subsequently, SWAP commutation rule~1 is applied to exchange the SWAP gate $g_4$ with the Hadamard gate $g_0$. 
    As a result, the solution $s_1$ is encoded as $s_1 = \{(g_3, g_4), (g_4, g_0)\}$. Solutions $s_2$ and $s_3$ are constructed in an analogous manner.
    \end{example}

    \item \textbf{Evaluation:} Each derived candidate solution is assessed using a predefined fitness function:
    \begin{equation}
    \label{eq:fit}
    \textit{fit}(s) = \alpha r^s_{\text{dpt}} + (1-\alpha) r^s_{\text{sub}},
    \end{equation}
    where $r_{\text{dpt}}$ and $r_{\text{sub}}$ represent the reductions in, respectively, depth and the number of CNOT subcircuits that can be extracted from the original circuit. The coefficient $\alpha$ ($0<\alpha<1$) achieves the trade-off between minimizing circuit depth and reducing the number of CNOT subcircuits. 
    Here, the CNOT subcircuits are identified using the same greedy sweeping strategy introduced in Sec.~\ref{sec:sweep}. (Line 2 in Alg.~\ref{alg:ga})

    \begin{example}
    The fitness values of the three candidate solutions $s_1$, $s_2$, and $s_3$ are computed as illustrated in Fig.~\ref{fig:ga_eg}(c)–(e). We set $\alpha = 0.9$.
    The original circuit has a depth of 12 and contains two CNOT subcircuits, namely $\{g_1, g_2, g_3, g_4\}$ and $\{g_6, g_7, g_8\}$.
    For solution $s_1$, the resulting circuit depth is reduced to 10, while the number of subcircuits remains 2, specifically $\{g_1, g_2, g_4\}$ and $\{g_3, g_6, g_7, g_8\}$. Hence, $fit(s_1) = 0.9 * (12 - 10) + 0.1 * (2-2) = 1.8$. For solution $s_2$, the circuit depth is 11 and all CNOT and SWAP gates are grouped into a single subcircuit, yielding $fit(s_2) = 0.9 * (12 - 11) + 0.1 * (2-1) = 1$. Similarly, solution $s_3$ has a depth of 11 with two subcircuits, resulting in $fit(s_3) = 0.9*(12-11)+0.1*(2-2)=0.9$. 
    \end{example}
    
    \item \textbf{Selection:} The top $n_{\text{species}}$ solutions with the highest fitness scores are selected from the population. (Line 4 in Alg.~\ref{alg:ga})
    \begin{example}
    In the example shown in Fig.~\ref{fig:ga_eg}, we select the three candidate solutions with the highest fitness values. We assume that the three solutions selected for the next generation are $s_1$, $s_2$, and $s_3$.
    \end{example}

    \item \textbf{Mutation:} Mutation is applied to $\alpha_{\mu}  n_{\text{species}}$ solutions, where $\alpha_{\mu}$ represents the mutation rate. For each selected solution, additional random commutation rules are introduced to boost the diversity of the population. (Lines 7 to 8 in Alg.~\ref{alg:ga})
    \begin{example}
    Assume that $\alpha_{\mu} = 0.7$. Accordingly, mutation is applied to $\alpha_{\mu} n_{\text{species}} = 0.7 \times 3 = 2.1 \approx 2$ candidate solutions, which are randomly selected. Here, we choose $s_2$ and $s_3$ for illustration. 
    For solution $s_2$, we apply SWAP commutation rule~4 to the SWAP gate $g_4$ and the CNOT gate $g_7$, and SWAP commutation rule~3 to the SWAP gate $g_6$ and the CNOT gate $g_8$. As a result, a new solution $s_4 = \{(g_4, g_5),(g_6, g_7),(g_4, g_7),(g_6, g_8)\}$ is generated, as shown in Fig.~\ref{fig:ga_eg}(f).
    Similarly, starting from solution $s_3$, we apply a mutation by exchanging the SWAP gate $g_6$ and the CNOT gate $g_8$, which yields the solution $s_5 = \{(g_6, g_7),(g_6,g_8)\}$ as illustrated in Fig.~\ref{fig:ga_eg}(g).
    \end{example}

    \item \textbf{Crossover:} We perform crossover operations $(1-\alpha_{\mu}) n_{\text{species}}$ times. In each crossove, two parent solutions are considered, with one being selected as the base. The commutation operations listed in the other solution are then applied, if valid, to the base. (Lines 9 to 13 in Alg.~\ref{alg:ga})

    \begin{example}
    In our example, $(1-\alpha_{\mu}) n_{\text{species}} = 0.3 \times 3 = 0.9 \approx 1$, and thus a single crossover operation is performed. Suppose that $s_1$ is selected as the base solution and $s_3$ is chosen as the second parent. The resulting offspring, denoted as $s_6$, is given by $\{(g_3, g_4), (g_4, g_0), (g_6, g_7)\}$ as illustrated in Fig.~\ref{fig:ga_eg}(h).
    Note that exchanging the SWAP gate $g_6$ and the CNOT gate $g_7$ is a valid commutation when applied to the circuit corresponding to $s_1$. 
    \end{example}

    \item \textbf{Replacement:} Following the crossover and mutation processes, the newly created solutions are re-evaluated with the same fitness function. The best $n_{\text{species}}$ solutions are then retained as the next generation. (Lines 14 to 19 in Alg.~\ref{alg:ga})

    \begin{example}
    For the six candidate solutions shown in Fig.~\ref{fig:ga_eg}(c)–(h), the corresponding fitness values are computed as $fit(s_1)=1.8$, $fit(s_2)=1$, $fit(s_3)=0.9$, $fit(s_4)=1$, $fit(s_5)=0.9$, and $fit(s_6)=1.8$.
    We then select the $n_{\text{species}}=3$ solutions with the highest fitness values, namely $s_1$, $s_6$, and $s_2$. In cases where multiple solutions share the same fitness value, one is selected uniformly at random. 
    \end{example}

    \item \textbf{Termination:} The evaluation, selection, mutation, and crossover process will be repeated until the algorithm either reaches the maximum number of iterations, $T_{\text{max}}$, or sees no improvement in the optimal fitness value for $T_{\text{idle}}$ iterations. Once either condition is met, we terminate the whole procedure and return the best solution. (Lines 6 to 21 in Alg.~\ref{alg:ga})

\end{itemize}

\begin{algorithm}
\caption{GABasedCommutation}
\label{alg:ga}
\begin{algorithmic}[1]
\REQUIRE A quantum circuit $\mathcal{C}$, population size $n_{\text{species}}$, trade-off coefficient $\alpha$, mutation rate $\alpha_{{\mu}}$, maximum iterations $T_{\text{max}}$, maximum idle times $T_{\text{idle}}$
\ENSURE A commutation sequence
\STATE Initialize population set $\mathcal{P}$ with $3\times n_{\text{species}}$ solutions through randomly applying commutation rules to $\mathcal{C}$ //Initialization
\STATE Evaluate each $s \in \mathcal{P}$ using $fit(s)$ in Equation~\ref{eq:fit}  //Evaluation 
\STATE $s^* \leftarrow \arg\max_{s\in \mathcal{P}}fit(s)$
\STATE Keep the best $n_{\text{species}}$ solutions in $\mathcal{P}$ //Selection
\STATE $i \leftarrow 1$
\WHILE{$ i \leq T_{\text{max}}$ or $s^*$ remains unchanged for $T_\text{idle}$ iterations}
\STATE $\mathcal{P}_M \leftarrow$ randomly select $\alpha_{\mu} * n_{\text{species}}$ solutions from $\mathcal{P}$ //Mutation (Line 7 to 8)
\STATE For each $s \in \mathcal{P}_M$, randomly apply commutation rules to $s$ and obtain a new solution
\STATE $\mathcal{P}_{\mathcal{C}} \leftarrow$ randomly select $(1-\alpha_{\mu}) * n_{\text{species}}$ solutions from $\mathcal{P}$ //Crossover (Lines 9 to 13)
\FOR{each $s$ in $\mathcal{P}_C$}
\STATE Randomly select another solution $s'$ from $\mathcal{P}$
\STATE Combine the solution $s$ with $s'$ and obtain a new solution 
\ENDFOR
\STATE $\mathcal{P}_{\text{new}} \leftarrow$ all solutions generated in this iteration via mutation ($\mathcal{P}_M$) and crossover ($\mathcal{P}_{\mathcal{C}}$)
\STATE $\mathcal{P} \leftarrow \mathcal{P} \cup \mathcal{P}_{\text{new}}$  // Add newly generated solutions to the population
\IF{$\max_{s\in \mathcal{P}}fit(s) > fit(s^*)$}
\STATE $s^* \leftarrow \arg\max_{s\in \mathcal{P}}fit(s)$
\ENDIF
\STATE Keep the best $n_{\text{species}}$ solutions in $\mathcal{P}$ //Replacement
\STATE $i \leftarrow i + 1$
\ENDWHILE
\STATE return $s^*$
\end{algorithmic}
\end{algorithm}

\subsection{Subcircuit sweeping}
\label{sec:sweep}

In our SSR optimization framework, the subcircuit sweeping phase serves to identify and extract CNOT subcircuits that can be further optimized via SAT-based rewriting. This stage involves three key tasks:
\begin{itemize}
    \item \emph{Scanning} and extracting subcircuits from the input quantum circuit, particularly those consisting of consecutive CNOT and SWAP gates.
    \item \emph{Evaluating} the optimization potential in depth reduction for each extracted subcircuit.
    \item  \emph{Selecting} the most promising subcircuits for rewriting.
\end{itemize}

\begin{figure}
\centering
\scalebox{1.2}{
\begin{tikzpicture}
\begin{yquant*}
    qubit {$\reg_{\idx}$} q[5];
    [name=left1]
    cnot q[0] | q[1];
    swap (q[1], q[2]);
    cnot q[1] | q[0];
    [name=left2]
    null q[2];
    [name=left3]
    cnot q[3] | q[2];
    [name=left5]
    cnot q[4] | q[3];
\end{yquant*}
\node[draw, dashed, fit=(left1) (left2)]{};
\node[draw, dashed, fit=(left3)(left5)]{};
\end{tikzpicture}
}
\caption{Subcircuit extraction up to $n_q=3$ qubits}
\label{fig:sub_extract_nq}
\end{figure}
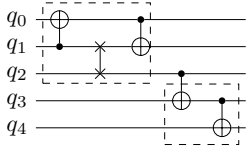

The scanning process scans throughout the input circuit from left to right, identifying and extracting all circuit blocks composed of contiguous CNOT and SWAP gates. Since the time overhead of the SAT solver for rewriting grows exponentially with the number of qubits in the subcircuit, we specify a maximum number of qubits $n_q$ for all extracted subcircuits to prevent excessive computational overhead. Fig.~\ref{fig:sub_extract_nq} shows an example of subcircuit extraction when $n_q = 3$.

Not all extracted subcircuits offer meaningful depth reductions when rewritten. To determine subcircuits with good potential, we introduce in the evaluation process a scoring function that estimates the potential depth improvement for each subcircuit: 
\begin{equation} 
\label{eq:score} \textit{score}(\mathcal{C}_\text{sub}) = d(\mathcal{C}) - d(\mathcal{C}/\mathcal{C}_\text{sub}) - d(\mathcal{C}_{\text{opt}}), 
\end{equation} where $\mathcal{C}$, $\mathcal{C}_{\text{sub}}$ and $\mathcal{C}_{\text{opt}}$ represent the input circuit, the subcircuit to be evaluated, and the depth-optimal circuit functionally equivalent to $\mathcal{C}_{\text{sub}}$, respectively,
and $d(\cdot)$ denotes the circuit depth. Here, $d(\mathcal{C})$ is the original circuit depth, $d(\mathcal{C}/\mathcal{C}_{sub})$ is the depth of the circuit after removing the subcircuit, and $d(\mathcal{C}_{\text{opt}})$ corresponds to the depth of its optimal replacement. Intuitively, this score reflects the potential depth reduction that can be achieved by substituting $\mathcal{C}_{\text{sub}}$ with its optimal implementation—a higher score indicates greater expected benefit.
After evaluating all extracted subcircuits, the top $n_t$ subcircuits with the highest scores are selected for rewriting. 

\begin{figure*}
\centering
\subfloat[Original circuit with three extracted CNOT subcircuits. The depth is 21 and the scores of each subcircuit are: 0, 1, 5]{
\begin{tikzpicture}
\begin{yquant}
    qubit {} q[4];
    barrier (q);
    cnot q[1] | q[0]; 
    cnot q[2] | q[1];
    cnot q[2] | q[3];
    cnot q[0] | q[1];
    barrier (q);
    box {$U$} q[0];
    barrier (q);
    swap (q[0], q[1]);
    cnot q[2] | q[1];
    cnot q[3] | q[2];
    cnot q[1] | q[2];
    cnot q[2] | q[3];
    cnot q[1] | q[2];
    barrier (q);
    box {$U$} q[1];
    barrier (q);
    swap (q[0], q[1]);
    cnot q[1] | q[0];
    cnot q[0] | q[1];
    cnot q[2] | q[1];
    cnot q[2] | q[3];
    barrier (q);
\end{yquant}
\end{tikzpicture}
}
\hfill
\subfloat[Circuit after simultaneous SAT-based rewriting of all subcircuits. The depth becomes 14.]{
\begin{tikzpicture}
\begin{yquant}
    qubit {} q[4];
    barrier (q);
    cnot q[1] | q[0]; 
    cnot q[2] | q[1];
    cnot q[2] | q[3];
    cnot q[0] | q[1];
    barrier (q);
    box {$U$} q[0];
    barrier (q);
    cnot q[0] | q[1];
    cnot q[1] | q[0];
    cnot q[2] | q[1];
    cnot q[0] | q[1];
    cnot q[3] | q[2];
    cnot q[1] | q[2];
    cnot q[2] | q[3];
    cnot q[1] | q[2];
    barrier (q);
    box {$U$} q[1];
    barrier (q);
    cnot q[1] | q[0];
    cnot q[2] | q[3];
    cnot q[2] | q[1];
    barrier (q);
\end{yquant}
\end{tikzpicture}
}
\hfill
\subfloat[Circuit obtained by rewriting only the third subcircuit (highest score)]{
\begin{tikzpicture}
\begin{yquant}
    qubit {} q[4];
    barrier (q);
    cnot q[1] | q[0]; 
    cnot q[2] | q[1];
    cnot q[2] | q[3];
    cnot q[0] | q[1];
    barrier (q);
    box {$U$} q[0];
    barrier (q);
    swap (q[0], q[1]);
    cnot q[2] | q[1];
    cnot q[3] | q[2];
    cnot q[1] | q[2];
    cnot q[2] | q[3];
    cnot q[1] | q[2];
    barrier (q);
    box {$U$} q[1];
    barrier (q);
    cnot q[1] | q[0];
    cnot q[2] | q[3];
    cnot q[2] | q[1];
    barrier (q);
\end{yquant}
\end{tikzpicture}
}
\hfill
\subfloat[Circuit after applying GA-based commutation to (c).]{
\begin{tikzpicture}
\begin{yquant}
    qubit {} q[4];
    barrier (q);
    cnot q[1] | q[0]; 
    cnot q[2] | q[1];
    cnot q[2] | q[3];
    cnot q[0] | q[1];
    swap (q[0], q[1]);
    barrier (q);
    box {$U$} q[1];
    barrier (q);
    cnot q[2] | q[1];
    cnot q[3] | q[2];
    cnot q[1] | q[2];
    cnot q[2] | q[3];
    cnot q[1] | q[2];
    barrier (q);
    box {$U$} q[1];
    barrier (q);
    cnot q[1] | q[0];
    cnot q[2] | q[3];
    cnot q[2] | q[1];
    barrier (q);
\end{yquant}
\end{tikzpicture}
}
\hfill
\subfloat[Circuit after the second-round rewriting of the first subcircuit. The depth becomes 13.]{
\begin{tikzpicture}
\begin{yquant}
    qubit {} q[4];
    barrier (q);
    cnot q[1] | q[0]; 
    cnot q[2] | q[1];
    cnot q[2] | q[3];
    cnot q[1] | q[0];
    cnot q[0] | q[1];
    barrier (q);
    box {$U$} q[1];
    barrier (q);
    cnot q[2] | q[1];
    cnot q[3] | q[2];
    cnot q[1] | q[2];
    cnot q[2] | q[3];
    cnot q[1] | q[2];
    barrier (q);
    box {$U$} q[1];
    barrier (q);
    cnot q[1] | q[0];
    cnot q[2] | q[3];
    cnot q[2] | q[1];
    barrier (q);
\end{yquant}
\end{tikzpicture}
}
\caption{An illustrative example demonstrating why rewriting only the top-ranked subcircuits in each iteration can lead to better global depth reduction than rewriting all subcircuits simultaneously. Rewriting all subcircuits at once (b) yields a suboptimal final depth compared to the iterative strategy that interleaves selective SAT-based rewriting with GA-based commutation (c–e).}
\label{fig:eg_subcirtop}
\end{figure*}

The rationale behind selecting only the top-scored subcircuits for SAT-based rewriting is that rewriting all extracted subcircuits simultaneously does not necessarily lead to better global optimization. Due to qubit overlap and depth interdependencies among subcircuits, aggressively optimizing every subcircuit may prematurely fix local structures and restrict further commutation opportunities.

To clarify why we do not rewrite all extracted subcircuits in each iteration, we provide an illustrative example shown in Fig.~\ref{fig:eg_subcirtop}.
Consider a 4-qubit circuit shown in Fig.~\ref{fig:eg_subcirtop}(a) from which three CNOT subcircuits are extracted after subcircuit sweeping. The original circuit has an overall depth of 21. The three subcircuits have depths 3, 8, and 7, respectively, while their corresponding depth-optimal implementations have depths 3, 7, and 2. According to our scoring function, which measures the potential depth reduction, their scores are therefore 0, 1, and 5.
If all three subcircuits are rewritten independently using SAT, the resulting depths of the subcircuits become 3, 7, and 2, leading to a total circuit depth of 14 (Fig.~\ref{fig:eg_subcirtop}(b) illustrates this resulting circuit). While this strategy already improves the circuit, it may prematurely eliminate SWAP structures that are essential for enabling further gate commutations, thereby limiting the ability of the subsequent genetic algorithm to explore beneficial reorderings and reducing the overall optimization potential.

Now consider the case where only the top-scoring subcircuit is rewritten in each iteration (i.e.,$n_t = 1$). In the first iteration, we rewrite the third subcircuit, which has the highest score (Fig.~\ref{fig:eg_subcirtop}(c)). After this rewriting, we apply GA-based commutation rules to the updated circuit. Suppose that a commutation between a single-qubit gate and a SWAP gate is applied, which changes the circuit structure and affects the subsequent subcircuit extraction (Fig.~\ref{fig:eg_subcirtop}(d)). As a result, the first subcircuit now has depth 6 with score 2, the second subcircuit has depth 5 with no further optimization potential (score 0), and the third subcircuit remains depth-optimal. 
In the second iteration, we rewrite the first subcircuit (Fig.~\ref{fig:eg_subcirtop}(e)). After this optimization, the overall circuit depth is reduced to 13, which is smaller than the depth obtained by rewriting all subcircuits independently in a single pass.

This example demonstrates that aggressively rewriting all subcircuits may prevent beneficial gate reordering opportunities across subcircuit boundaries. By prioritizing subcircuits with the highest optimization potential and interleaving SAT-based rewriting with GA-based commutation, the algorithm can expose new optimization opportunities in later iterations, leading to better global depth reduction. Therefore, the selective rewriting strategy in SSR is designed to balance exact local optimization and global structural flexibility, leading to improved overall circuit depth in practice.

Note that it is computationally expensive to calculate the exact value of $d(\mathcal{C}_{\text{opt}})$ because it requires the exact synthesis of linear transformations, which is NP-hard.
To mitigate this issue, we employ a supervised machine learning approach to approximately calculate $d(\mathcal{C}_{\text{opt}})$ of each subcircuit. Some important technical details are listed as follows.

\paragraph{Training dataset generation}

To construct the training dataset for the ANN, we exploit the fact that any CNOT–SWAP circuit can be uniquely represented by an invertible Boolean matrix over $\mathbb{F}_2$~\cite{markov2008optimal,wu2023optimization}.
For a fixed number of qubits and a given architecture graph (AG), we first randomly generate a large set of CNOT circuits whose two-qubit gates strictly respect the hardware connectivity constraints imposed by the AG.
Each generated circuit is then mapped to its corresponding invertible Boolean matrix, which serves as the ANN input feature, where each matrix entry is treated as an independent input variable.

To obtain the training labels, we compute the depth-optimal CNOT implementation for each matrix using the SAT-based formulation described in Sec.~\ref{sec:cnf_form}, and record the optimal depth as the ground-truth label.
Since the achievable optimal depth depends on the underlying connectivity, we construct a separate training dataset for each AG.

For experimental evaluation, the test circuits are generated independently from the training set, ensuring that no Boolean matrix or circuit instance used during training appears in the evaluation phase.

\paragraph{Model selection and learning}

The Artificial Neural Network (ANN), specifically a Multi-layer Perceptron (MLP), is utilized as the prediction model. MLPs are widely used due to their ability to learn intricate and non-linear relationships between input and output variables across multiple layers of interconnected neurons. To avoid overestimation, the loss function used for the learning process is a penalized Mean Squared Error (MSE):

\begin{equation}
\label{eq:loss}
\frac{1}{n} \sum_{i=1}^n\left\{\begin{array}{ll}
\left(y_i-\hat{y}_i\right)^2, & \text { if } \hat{y}_i \leq y_i \\
(1+\beta) \cdot\left(y_i-\hat{y}_i\right)^2, & \text { if } \hat{y}_i>y_i
\end{array}\right.
\end{equation}
where $y_i$ is the true optimal depth of the $i$-th sample, $\hat{y}_i$ is the predicted depth of the $i$-th sample, $n$ is the total number of samples in the training dataset, and $\beta > 0$ is the penalty coefficient to control the weight of overestimation in the loss function.  
The designed loss function penalizes overestimation more heavily than underestimation, which is crucial because overestimating the depth may bring significant computation overhead during the following SAT-based rewriting procedure as will be explained later.

\paragraph{Optimal depth prediction}
Once trained, ANN will be used to predict the optimal depth of each input subcircuit. Given an input circuit, we first identify the architecture graph and select the corresponding model. Then the Boolean matrix representation of the circuit will be extracted and fed into the specific model to predict the optimal depth.

\subsection{SAT-based subcircuit rewriting}

\label{sec:disable}
In the rewriting phase, each selected subcircuit is replaced with a functionally equivalent circuit that achieves the minimal possible depth while maintaining compliance with hardware connectivity constraints. To obtain the depth-optimal equivalent circuit, we formulate such an exact synthesis problem of CNOT circuits as an SAT problem and utilize an off-the-shelf SAT solver to address it. The basic CNF encoding scheme used in this process is based on a prior SAT-based synthesis approach ~\cite{chen2022optimizing}, also shown in Sec.~\ref{sec:cnf_form}. In particular, two key modifications are also introduced to improve the efficiency and effectiveness in reducing the overall circuit depth, including ANN-assisted SAT invoking and gate position constraining {without sacrificing optimality.} 

\paragraph{Gate position constraining}
Although for each CNOT subcircuit, the depth-optimal circuit can be found via the original CNF encoding scheme in~\cite{chen2022optimizing}, this local optimization at the subcircuit level does not always translate to a reduction in the depth of the overall quantum circuit. This issue arises from the intricate interdependencies between subcircuits and their interactions with the remaining components of the circuit.

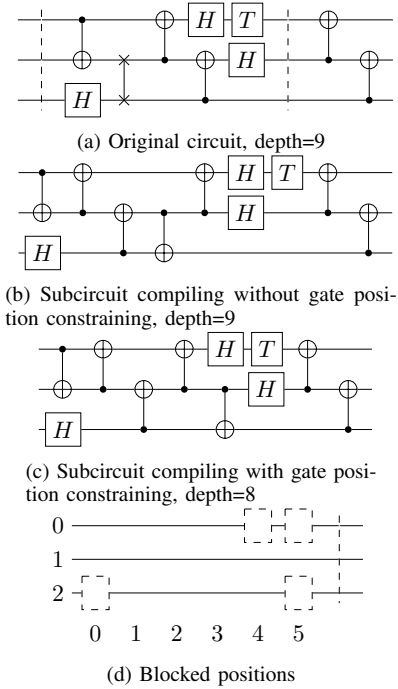
\begin{figure}
\centering
\scalebox{1}{
\subfloat[Original circuit, depth=9, the H gate located in (7,2) can be shifted to the location (8, 2)]{
\begin{tikzpicture}
\begin{yquant}
    qubit { $\idx$} q[3];
    nobit depth;
    barrier (q);
    h q[2];
    cnot q[1] | q[0]; 
    cnot q[1] | q[2];
    cnot q[2] | q[1];
    cnot q[1] | q[2];
    cnot q[0] | q[1];
    h q[0];
    cnot q[1] | q[2];
    box {$T$} q[0];
    h q[1];
    h q[2];
    barrier (q);
    [dashed]box {} q[2];
    cnot q[0] | q[1];
    cnot q[1] | q[2];
    
    [draw=none] null depth;
    [draw=none] box {$1$} depth;
    [draw=none] box {$2$} depth;
    [draw=none] box {$3$} depth;
    [draw=none] box {$4$} depth;
    [draw=none] box {$5$} depth;
    [draw=none] box {$6$} depth; 
    [draw=none] box {$7$} depth;
    [draw=none] null depth;
    [draw=none] box {$8$} depth;
    [draw=none] box {$9$} depth;
\end{yquant}
\end{tikzpicture}
}
}
\hfill
\scalebox{1}{
\subfloat[Subcircuit compiling without gate position constraining, depth=9]{
\begin{tikzpicture}
\begin{yquant}
    qubit { $\idx$} q[3];
    nobit depth;
    h q[2];
    cnot q[1] | q[0];
    cnot q[0] | q[1];
    cnot q[1] | q[2];
    cnot q[2] | q[1];
    cnot q[0] | q[1];
    h q[2];
    h q[0,1];
    box {$T$} q[0];
    cnot q[0] | q[1];
    cnot q[1] | q[2];

    [draw=none] box {$1$} depth;
    [draw=none] box {$2$} depth;
    [draw=none] box {$3$} depth;
    [draw=none] box {$4$} depth;
    [draw=none] box {$5$} depth;
    [draw=none] box {$6$} depth;
    [draw=none] box {$7$} depth;
    [draw=none] box {$8$} depth;
    [draw=none] box {$9$} depth;
\end{yquant}
\end{tikzpicture}
}
}
\vfill
\scalebox{1}{
\subfloat[Subcircuit compiling with gate position constraining, depth=8]{
\begin{tikzpicture}
\begin{yquant}
    qubit { $\idx$} q[3];
    nobit depth;
    h q[2];
    cnot q[1] | q[0];
    cnot q[0] | q[1];
    cnot q[1] | q[2];
    cnot q[0] | q[1];
    h q[0];
    cnot q[2] | q[1];
    box {$T$} q[0];
    h q[1];
    h q[2];
    cnot q[0] | q[1];
    cnot q[1] | q[2];

    [draw=none] box {$1$} depth;
    [draw=none] box {$2$} depth;
    [draw=none] box {$3$} depth;
    [draw=none] box {$4$} depth;
    [draw=none] box {$5$} depth;
    [draw=none] box {$6$} depth;
    [draw=none] box {$7$} depth;
    [draw=none] box {$8$} depth;
\end{yquant}
\end{tikzpicture}
}
}
\hfill
\scalebox{1}{
\subfloat[Blocked positions (1,2), (5,0), (6,0) and (6,1)]{
\begin{tikzpicture}
\begin{yquant*}
    qubit { $\idx$} q[3];
    nobit depth;
    [dashed]box {} q[2];
    null q[1];
    null q[1];
    null q[1];
    null q[1];
    null q[1];
    [dashed]box {} q[1];
    null q[0];
    null q[0];
    null q[0];
    null q[0];
    [dashed]box {} q[0];
    [dashed]box {} q[0];
    barrier (q);
    [draw=none] box {$1$} depth;
    [draw=none] box {$2$} depth;
    [draw=none] box {$3$} depth;
    [draw=none] box {$4$} depth;
    [draw=none] box {$5$} depth;
    [draw=none] box {$6$} depth;

\end{yquant*}
\end{tikzpicture}
}
}
\caption{Gate position constraining}
\label{fig:dis_eg}
\end{figure}

An illustrative example of this phenomenon is depicted in Fig.~\ref{fig:dis_eg}. 
After rewriting the subcircuit using the SAT solver (cf. Fig.~\ref{fig:dis_eg}(b)), the depth of the subcircuit decreased from 6 to 5. However, the overall circuit depth remains unchanged at 9. This is because the SAT solver positioned the CNOT gates in locations that are critical for determining the circuit's depth.

To address this challenge, 
we introduce additional constraints to the CNF encoding scheme in~\cite{chen2022optimizing} to account for the ``gate position constraining''. 
Specifically, these constraints prevent the placement of CNOT gates in blocked positions that are already occupied by some gates from the QCT circuit. Consequently, the SAT solver is prompted to seek an alternative solution that adheres to these constraints, leading to a more effective reduction in overall circuit depth. 
Fig.~\ref{fig:dis_eg}(c) demonstrates the effectiveness of this strategy: by blocking the position shown in Fig.~\ref{fig:dis_eg}(d), we reduce the total circuit depth from 9 to 8 through rewriting. 

To apply the gate position constraining in a systematic manner, we next explain how the blocked positions are identified.
Each gate in the circuit will be associated with the space--time coordinates $(d,q)$ (two-qubit gate has two coordinates), indicating that the gate acts on qubit(s) $q$ at depth $d$.
For a given extracted CNOT subcircuit $\mathcal{C}_i$, we denote
$d(\mathcal{C}_i)_{\min} = \min_{(d,q)\in \mathcal{C}_i}{d}.$
If this subcircuit is followed by another CNOT subcircuit $\mathcal{C}_{i+1}$, then the depth range $[d(\mathcal{C}_i)_{\min},\, d(\mathcal{C}_{i+1})_{\min}-1]$ and the occupied qubit set define the valid window in which the current subcircuit can be rewritten.

Within this valid window, blocked positions are determined by inspecting all gates that are not part of the current CNOT subcircuit. Once these gates are identified, positions matching their coordinates are marked as “blocked” for the subsequent SAT encoding. Consider the original circuit shown in Fig.~\ref{fig:dis_eg}(a), where the extracted CNOT subcircuit, say $\mathcal{C}_{1}$, lies between the dotted lines. It can be found that $d(\mathcal{C}_{2})_{\min}-1 = 7$ where $\mathcal{C}_{2}$ denotes the CNOT subcircuit next to $\mathcal{C}_{1}$. Therefore
the valid window of $\mathcal{C}_{1}$ lies between depth 1 to 7, and the H and T gates at position (1,2) (6,0), (7,0) and (7,1) are treated as blocked. Since position (8,2) is vacant, the gate originally at (7,2) is shifted to (8,2) to free up additional room for the upcoming rewriting step. Furthermore, the blocked positions on the right side, i.e., (6,0),(7,0),(7,1), may be shifted left, depending on the target depth used in the SAT-based rewriting. Suppose current target depth is 6, then these three right-side blocked positions, initially at (6,0), (7,0), and (7,1), will be relocated to (5,0), (6,0), and (5,1), respectively, to fit this maximum permitted depth (cf. Fig.~\ref{fig:dis_eg}(d)).

We introduce additional constraints, referred to as gate position constraints,
which are derived from a systematic identification of blocked positions
as described in Alg.~\ref{alg:block}.

\begin{algorithm}
\caption{BlockedPositionIdentification}
\label{alg:block}
\begin{algorithmic}[1]
\REQUIRE Quantum circuit $\mathcal{C}_{\text{ori}}$, CNOT subcircuit $\mathcal{C}$, target SAT depth $D$
\ENSURE Set of blocked positions $P$
\STATE $P \leftarrow \emptyset$
\STATE $\mathcal{C'} \leftarrow $ the CNOT subcircuit next to $\mathcal{C}$ 

\COMMENT{Identify left-side blocked positions}

\FOR{$d = d(\mathcal{C})_{\min}$ to $d(\mathcal{C'})_{\min}-1$}
    \FORALL{gates $g$ at depth $d$ and on the left side of $\mathcal{C}$}
        \IF{$g \notin \mathcal{C}$}
            \FORALL{qubits $q$ acted on by $g$ and occupied by $\mathcal{C}$}
                \STATE $P \leftarrow P \cup \{(d,q)\}$
            \ENDFOR
        \ENDIF
    \ENDFOR
\ENDFOR

\COMMENT{Identify right-side blocked positions with relocation}
\STATE Initialize $R[q] \leftarrow 0$ for all qubits $q$ {in $\mathcal{C}$}
\FOR{$d = d(\mathcal{C'})_{\min}-1$ to $d(\mathcal{C})_{\min}$}
    \FORALL{gates $g$ at depth $d$ and on the right side of $\mathcal{C}$}
        \IF{$g \notin \mathcal{C}$}
            \IF{$g$ can be right-shifted to a position with depth larger than $d(\mathcal{C'})_{\min}-1$}
                \STATE continue
            \ELSE
                \FORALL{qubits $q$ acted on by $g$ and occupied by $\mathcal{C}$}
                    \STATE $R[q] \leftarrow R[q] + 1$
                \ENDFOR
            \ENDIF
        \ENDIF
    \ENDFOR
\ENDFOR

\COMMENT{Assign right-side blocked positions {based on} target depth $D$}
\FORALL{qubit $q$ {in $\mathcal{C}$}}
    \IF{$R[q]=0$}
    \STATE continue
    \ELSE
    \FOR{$i = 0$ to $R[q]-1$}
        \STATE $P \leftarrow P \cup \{(D-i, q)\}$
    \ENDFOR
    \ENDIF
\ENDFOR
\STATE \textbf{return} $P$
\end{algorithmic}
\end{algorithm}

A blocked position can be represented as $(d,q)$, indicating that no CNOT gate involving qubit $q$ should be placed at depth $d$. We formalize this constraint as follows: 
for each blocked position $(d, q)$,
$$
(d,q) \Longrightarrow \bigvee_{c \in \delta(q)} g^d_{c\to q} \vee g^d_{q \to c} = 0,
$$
where  $\delta(q)$ is the set of qubits adjacent to $q$, and $g^d_{c \to q}$  indicates that a CNOT between qubits $c$ and $t$ took place at depth $d$.

By encoding blocked positions as hard constraints in the CNF formulation,
the SAT solver is restricted from assigning CNOT gates to depth-critical
locations, thereby reducing the search space to solutions that are more
likely to yield a global depth reduction.

It is important to explicitly clarify the modeling scope of our SAT formulation. As currently described, the SAT encoding strictly optimizes CNOT-only subcircuits, meaning that single-qubit gates are not included as variables within the search space and their placements remain fixed. 
Under this restricted scope, we introduce a gate position constraining mechanism that enforces the rewritten CNOT subcircuit to occupy the same space‑time window as the original subcircuit. This mechanism is essential for maintaining global depth consistency when a local CNOT substitution is performed inside a fixed schedule of surrounding gates.

Excluding single-qubit gates is both a deliberate design choice and a computational necessity. From a design perspective, the SSR framework leverages the scalability of subcircuit rewriting to handle large-scale circuits, a strategy necessitated by the inherent complexity of global SAT-based synthesis~\cite{haaswijk2019sat}. Regarding computational justifications,  SSR specifically targets routing-induced CNOT-SWAP overhead, and performing exact SAT synthesis for general unitaries would be prohibitively expensive~\cite{gouzien2025provably}. Furthermore, SAT solvers operate on strict Boolean logic and are fundamentally ill-suited for encoding continuous numerical variables, such as arbitrary rotation angles. The computational complexity of such universal synthesis is formidable, with classical approaches exhibiting doubly exponential time in the worst case~\cite{zak2025reducing}. Even when restricted to sub-universal gate sets like Clifford groups, exact SAT-based depth optimization currently hits a strict scalability wall beyond 6 qubits~\cite{schneider2023sat, shaik2025cnot, peham2023depth}.

We emphasize that the necessity of gate position constraining is distinct from the decision to exclude single‑qubit gates. Gate position constraining arises directly from the subcircuit‑based decomposition of the optimization problem, independent of whether single‑qubit gates are modeled as variables. When rewriting a local CNOT subcircuit within a fixed global schedule of single‑qubit gates, the replaced subcircuit must respect the space‑time boundaries imposed by surrounding gates; otherwise, even a locally optimal CNOT realization could shift its logical positions and inadvertently delay depth‑critical operations outside the window. Thus, this mechanism serves as an essential boundary manager for global depth consistency under any subcircuit rewriting scheme.

The exclusion of single‑qubit gates is justified separately on pragmatic grounds of scalability and feasibility, not by any claim that the CNOT‑only model preserves global optimality. We acknowledge that the resulting solution is not guaranteed to remain locally optimal after reinserting single‑qubit gates, and that the scenario where a strictly CNOT‑optimal configuration remains optimal under the full gate set relies on assumptions that do not hold in general. Instead, our modeling choice is a deliberate engineering trade‑off: we sacrifice exact optimality for the ability to optimize CNOT depth and gate count on circuits up to two orders of magnitude larger than what full SAT synthesis can handle.

\paragraph{ANN-assisted SAT invoking}
In the original implementation \cite{chen2022optimizing}, the SAT solver is invoked in a trial-and-error manner, indicating that for each possible depth $d$, the solver will be called and (if possible) output an equivalent CNOT circuit with that depth. This process begins at $d=1$ and is iteratively repeated until an optimal depth is reached. While this method is straightforward, it can be quite time-consuming, especially when the optimal depth is large.

To mitigate the time consumption associated with failed trials, we use the trained ANN model in Sec.~\ref{sec:sweep} to predict the optimal depth of the input subcircuit, denoted as $d_\text{pred}$. Instead of starting from $d=1$, we use $d_\text{pred}$ as the target depth for the initial trial (Line 2 in Alg.~\ref{alg:sat_total}).

The process then proceeds as follows:
\begin{itemize}
    \item Successful Case ($d = d_\text{pred}$): If the SAT solver successfully constructs a circuit with depth $d = d_\text{pred}$, we attempt to determine whether a circuit with depth  $d - 1 $ exists. This is done iteratively by decrementing $d$ and invoking the SAT solver, continuing until no valid solution is found (Lines 4 to 9 in Alg.~\ref{alg:sat_total}). The smallest valid depth is then chosen as the optimal depth.  
    \item Failure Case ($d = d_\text{pred}$ Fails): If the SAT solver fails to construct a circuit at $d_\text{pred}$, we increment $d$ and invoke the solver again (Lines 11–15 in Alg.~\ref{alg:sat_total}). This process is repeated until a feasible solution is obtained.
\end{itemize}

To ensure that the prediction does not significantly overestimate the depth, the ANN training incorporates a penalty function in the loss calculation (Eq.~\ref{eq:loss}). This penalty is designed to discourage overestimation, meaning that in most cases, $d_\text{pred}$ is an underestimate. As a result, our search process typically proceeds towards an upward direction, leading to a faster termination compared to the original trial-and-error method. 

By leveraging the ANN for depth prediction, our approach significantly reduces the computational overhead associated with SAT-based subcircuit optimization. Experimental results in Sec.~\ref{sec:core_tech_eva} demonstrate this improvement.

\begin{algorithm}
\caption{SATBasedRewriting}
\label{alg:sat_total}
\begin{algorithmic}[1]
\REQUIRE The input circuit $\mathcal{C}_{\text{ori}}$, CNOT subcircuit $\mathcal{C}$, predict target depth $d_\text{pred}$ 
\ENSURE Compiled subcircuit $\mathcal{C}_{\text{opt}}$
\STATE $P \leftarrow$ BlockedPositionIdentification($\mathcal{C}_{\text{ori}}, \mathcal{C}$, $d_{\text{pred}}$)
\STATE $\mathcal{C}_{\text{opt}} \leftarrow$ SATSolver($\mathcal{C}$, $d_\text{pred}$, $P$). 
\STATE Let $d \leftarrow d_\text{pred}$
\IF{$C_{\text{opt}}$ is not empty}
\REPEAT
\STATE $d \leftarrow d - 1$
\STATE $C_{\text{opt}}' \leftarrow $ SATSolver($\mathcal{C}$, $d$, $P$).
\STATE $\mathcal{C}_{\text{opt}} \leftarrow \mathcal{C}_{\text{opt}}'$ if $\mathcal{C}_{\text{opt}}'$ is not empty
\UNTIL{$\mathcal{C}_{\text{opt}}'$ is empty}
\ELSE
\REPEAT
\STATE $d \leftarrow d + 1$
\STATE $\mathcal{C}_{\text{opt}}' \leftarrow$ SATSolver($\mathcal{C}$, $d$, $P$).
\STATE $\mathcal{C}_{\text{opt}} \leftarrow \mathcal{C}_{\text{opt}}'$ if $\mathcal{C}_{\text{opt}}'$ is not empty
\UNTIL{$\mathcal{C}_{\text{opt}}'$ is not empty }
\ENDIF
\STATE Return $\mathcal{C}_{\text{opt}}$
\end{algorithmic}
\end{algorithm}

\section{Evaluation}
In this section, we conduct comprehensive evaluations across multiple quantum circuit benchmarks and hardware architectures, confirming the effectiveness of the proposed SSR optimization framework for QCT.

\subsection{Experimental setup}
We implement the proposed algorithm in Python\footnote{The code can be provided upon request.}. All experiments are conducted on a MacBook Pro featuring a 2.3
GHz Intel Core i5 processor and 16 GB memory. The benchmark circuits include RevLib~\cite{wille2008revlib}, quantum circuits extracted from Qiskit\footnote{https://docs.quantum.ibm.com/api/qiskit/circuit\_library}, which consist of IQP, QFT, Quantum Volume, Linear Pauli Rotation, Two Local circuits, and random circuits each of which contains  50\% randomly placed single-qubit gates and 50\% randomly placed CNOT gates. The tested AGs are Grid 5x4 with 20 qubits, Google Sycamore~\cite{arute2019quantum} and IBM Rochester with 53 qubits, and IBM Heron with 156 qubits.
The value of parameter $n_q$ (the maximum number of qubits in a sub-circuit) is set to 5, and $n_t$ (the ratio of highest scores) is set to 50\% of the subcircuits. 
The reason for choosing the value of $n_q$ is that as $n_q$ increases, the running time of SAT-Solver will increase exponentially. In order to balance efficiency and time, $n_q = 5$ is finally chosen.
For the ratio of highest scores $n_t$, we conducted experiments with RevLib on Grid 5x4, which is shown in Fig.~\ref{fig:compileRatio}. According to the result, we set $n_t$ to 50\%.
\begin{figure}
\centering
    \pgfplotsset{
        index_axis/.style={
            xlabel={The ratio of highest scores of the compiled nodes},
            ylabel={Depth improvement},
            xtick={0.1, 0.2, 0.3, 0.4, 0.5, 0.6, 0.7, 0.8, 0.9, 1},
            xticklabel style={
                /pgf/number format/fixed,
                /pgf/number format/precision=3,
                rotate=45,
                anchor=east
            },
            width = 12cm,
            height = 8cm,
            grid=both,
            mark repeat=1,
            xlabel style = {font =\Large},
            ylabel style = {font =\Large},        
        }
        }
    \scalebox{0.65}{
    \begin{tikzpicture}
\begin{axis}[index_axis ]

\addplot [SSRStyle]
table {%
0.1     0.286566009
0.2 	0.290882695
0.3 	0.291792489
0.4 	0.29527681
0.5 	0.296767325
0.6 	0.292605497
0.7 	0.288908246
0.8 	0.281881533
0.9 	0.277971351
1   	0.27841657
};

\end{axis}
\end{tikzpicture}
    }
    \caption{Evaluations for various $n_t$ for RevLib on Grid 5x4. The depth improvement is defined as $\frac{D_{\text{ori}}-D_{\text{opt}}}{D_{\text{ori}}}$, where $D_{\text{ori}}$ represents the original depth before optimization and $D_{\text{opt}}$ represents the depth after optimization.}
    \label{fig:compileRatio}
\end{figure}

For the ANN models, we employ the ReLU activation function due to its simplicity and effectiveness in training deep neural networks. The models are trained using the L-BFGS optimizer~\cite{liu1989limited}, which is particularly suitable for small- to medium-sized datasets and enables fast convergence.

The input to the ANN is a fixed-length feature vector of dimension 25. Specifically, each CNOT subcircuit acting on at most $n_q = 5$ qubits is represented by its corresponding $5 \times 5$ invertible Boolean matrix over $\mathbb{F}_2$, which captures the linear transformation implemented by the CNOT circuit. The matrix is flattened in row-major order and is used directly as the input of ANN. The output layer has dimension 1 and predicts the estimated optimal circuit depth of the corresponding CNOT subcircuit under a specific architecture graph (AG).

We employ four fully connected hidden layers with sizes 200, 150, 100, and 50, respectively. Because the optimal depth of a CNOT subcircuit is tied to the specific hardware connectivity, we independently train multiple ANNs for all sub-AGs that appear in the experiment and some common used sub-AGs. In total, 13 such AGs are identified, leading to 13 separately trained ANN models. For training, we use a modified loss function (Eq.~\ref{eq:loss}) that penalizes overestimation of the depth. This approach effectively prevents depth overestimation and thereby avoids unnecessary SAT invocations.

\begin{table}[!ht]
    \centering
    \caption{Performance metrics of all ANN models.
    Accuracy ($\pm$ X) means the ratio of predictions on test sets satisfying $|D_{\text{pred}} - D_{\text{opt}}| \le X$ where $D_\text{pred}$ and $D_{\text{opt}}$ represent the ANN predicted and the optimal depths, respectively.
    Overestimate reports the ratio of over-predicted cases, and MSE denotes the mean squared error between the predicted and optimal depths.}
    \label{tab:ANN_AG}
    \begin{tabular}{|c|c|c|c|c|c|c|}
    \hline
        AG No. & Accuracy ($\pm$0) & Accuracy ($\pm$1) & Accuracy ($\pm$2) & Overestimate & MSE & Training Time(s) \\ \hline
        1 & 50.2\% & 72.8\% & 96.6\%
        & 27.6\% & 0.898 & 195.93 \\ \hline
        2 & 65.4\% & 87.2\% & 99.4\% & 15\% & 0.434 & 169.5 \\ \hline
        3 & 59.4\% & 86.4\% & 99.6\%& 15.2\% & 0.464 & 140.3 \\ \hline
        4 & 62\% & 86.2\% & 99.2\% & 12.6\% & 0.452 & 119.6 \\ \hline
        5 & 63.2\% & 87.2\% & 99.6\% & 13.2\% & 0.452 & 130.3 \\ \hline
        6 & 60\%  & 87.6\% & 99.8\% & 14\% & 0.466 & 159.0 \\ \hline
        7 & 67.2\% & 91.6\% & 100\% & 8.8\% & 0.394 & 160.0 \\ \hline
        8 & 70.4\%  & 91\% & 99.4\% & 9.2\% & 0.35 & 141.1 \\ \hline
        9 & 72.2\% & 94.8\% & 100\% & 10.6\% & 0.29 & 121.1 \\ \hline
        10 & 66.4\% & 90.8\% & 99.4\% & 14.2\% & 0.418 & 159.5 \\ \hline
        11 & 58.8\% & 83.4\% & 98.2\% & 15.4\% & 0.604 & 132.1 \\ \hline
        12 & 68\% & 86.8\% & 99\% & 10\% & 0.338 & 144.0 \\ \hline
        13 & 61.8\% & 91.8\% & 100\% & 11.6\% & 0.424 & 171.2 \\ \hline
    \end{tabular}
    
\end{table}

Tab.~\ref{tab:ANN_AG} details performance metrics for all 13 ANN models. Overall, the models achieve moderate exact prediction accuracy, with Accuracy ($\pm$0) values typically ranging from approximately 50.2\% to 72.2\%. When allowing small prediction tolerances, the performance improves substantially: Accuracy ($\pm$1) consistently exceeds 83\% across all AGs, and Accuracy ($\pm$2) is close to or equal to 100\% in most cases. This indicates that the majority of prediction errors are bounded within a small depth deviation. More importantly, the average probability that the predicted depth exceeds the true optimal depth is only 13.65\%, indicating that the ANN outputs are biased toward underestimation, as intended. Across the 13 models, the mean squared error (MSE) ranges from 0.29 to 0.898, with an average MSE of approximately 0.46.

For GA, the population size $n_{\text{species}}$ is 10, the trade-off coefficient $\alpha$ in Eq.~\ref{eq:fit} is 0.9, the mutation rate $\alpha_{\mu}$ is 0.4, the maximum iterations $T_{\text{max}}$ is 50, and the maximum idle times $T_{\text{idle}}$ is 15.

\subsection{Effectiveness of each core technique}
\label{sec:core_tech_eva}

We first conduct experiments to verify the effectiveness of the three core techniques: gate position constraining ( Sec.~\ref{sec:disable}), genetic algorithm (Sec.~\ref{sec:swapping_ga}), and ANN depth prediction(Sec.~\ref{sec:sweep}), on the overall performance. To this end, we choose the circuits from RevLib and compiled them with the state-of-the-art QCT approach SABRE, which has already been integrated into Qiskit, using a hypothetical 3x3 grid as the underlying AG. Moreover, we implement three different versions of SSR, each corresponding to a simplified version created by removing some of the core techniques, as shown in Tab.~\ref{tab:four_version}.

Specifically, \textbf{SSR1} serves as the baseline implementation of our framework. It repeatedly (i) scans the circuit to extract consecutive CNOT subcircuits, (ii) evaluates each subcircuit using the score defined in Eq.~\ref{eq:score}, where the optimal depth is computed exactly by an SAT solver, and (iii) selects the top $n_t$ subcircuits for SAT-based rewriting without any gate position constraining. This process is iterated until no further depth reduction is observed. \textbf{SSR2} extends SSR1 by incorporating the proposed gate position constraining mechanism to prevent locally depth-optimal rewritings.
\textbf{SSR3} further augments SSR2 with the GA-based commutation search, enabling the exploration of additional gate reordering opportunities and exposing new CNOT subcircuits for further rewriting.
Finally, \textbf{SSR} (full version) enhances SSR3 with ANN-assisted depth prediction, which is used to accelerate SAT invocations and guide subcircuit prioritization.

\begin{table}[h]
    \centering
\caption{Three simplified versions of the SSR method}
    \begin{tabular}{|m{1cm}<{\centering}|m{1cm}<{\centering}|m{1cm}<{\centering}|m{1cm}<{\centering}|m{1cm}<{\centering}}
    \hline
        Version & Gate Position & GA & ANN  \\ \hline
        SSR & \ding{51} & \ding{51} & \ding{51} \\ \hline
        SSR1&  $\times$ & $\times$ & $\times$  \\ \hline
        SSR2&  \ding{51} & $\times$ & $\times$  \\ \hline
        SSR3&  \ding{51} & \ding{51} & $\times$  \\ \hline
    \end{tabular}
\label{tab:four_version}
\end{table}

\paragraph{Gate position constraining for SAT solver}
To demonstrate the effectiveness of introducing the gate position constraints into the SAT solver, we compare SSR1 with SSR2. The results are shown in Tab.~\ref{tab:disable}. It is easy to see that through introducing such constraints, the overall depth can be reduced and this reduction effect increases with the original circuit depth.

\begin{table}[h]
    \centering
    \caption{Evaluation for various simplified versions of SSR where $d(X)$ is the depth of the output circuit obtained by algorithm X.}
    \begin{tabular}{c|c|c|c|c|c}
    \hline
        Name & d(ori) &  d(SSR1) & d(SSR2) & d(SSR3) & d(SSR) \\ \hline
        4mod5-bdd\_287 & 89 &  78  & 76   & 68  & 68\\ \hline
        rd53\_135 & 339 &  271  & 269  & 249 & 244 \\ \hline
        ham7\_104 & 418 &  326  & 320  & 267 &  273 \\ \hline
        rd53\_131 & 568 &  444  & 437  & 398  & 392 \\ \hline
        rd53\_133 & 752 &  572  & 567  & 508  & 512 \\ \hline
        ex2\_227 & 756 &  632 & 628  & 557 &  557 \\ \hline
        majority\_239 & 806 &  663  & 659 & 541  &  539 \\ \hline
        rd53\_130 & 1271  &  1051  & 1044  & 888 &  890 \\ \hline
        f2\_232 & 1526 &  1192   & 1184  & 1024  &  1027 \\ \hline
        rd53\_251 & 1588 &  1287 & 1262 & 1195  &  1192 \\ \hline 
        Summation & 8113 & 6516 & 6446  & 5695  & 5694\\ \hline
    \end{tabular}
\label{tab:disable}
\end{table}

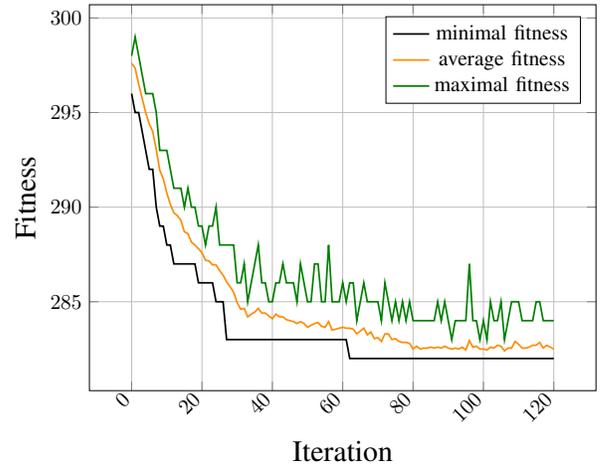
\begin{figure}
\centering
    \pgfplotsset{
        ga_axis/.style={
            xlabel={Iteration},
            ylabel={Fitness},
            xtick={0, 20, 40, 60, 80, 100, 120},
            xticklabel style={
                /pgf/number format/fixed,
                /pgf/number format/precision=3,
                rotate=45,
                anchor=east
            },
            width = 10cm,
            height = 8cm,
            grid=both,
            mark repeat=1,
            xlabel style = {font =\Large},
            ylabel style = {font =\Large},        
        }
        }
    \scalebox{0.8}{
\begin{tikzpicture}
\begin{axis}[ga_axis, legend entries={minimal fitness,average fitness, maximal fitness},
legend pos=north east,
]
\addplot [SSRStyle]
table {
0 296
1 295
2 295
3 294
4 293
5 292
6 292
7 290
8 289
9 289
10 288
11 288
12 287
13 287
14 287
15 287
16 287
17 287
18 287
19 286
20 286
21 286
22 286
23 286
24 285
25 285
26 285
27 283
28 283
29 283
30 283
31 283
32 283
33 283
34 283
35 283
36 283
37 283
38 283
39 283
40 283
41 283
42 283
43 283
44 283
45 283
46 283
47 283
48 283
49 283
50 283
51 283 
52 283
53 283
54 283
55 283
56 283
57 283
58 283
59 283
60 283
61 283
62 282
63 282
64 282
65 282
66 282
67 282
68 282
69 282
70 282
71 282
72 282
73 282
74 282
75 282
76 282
77 282
78 282
79 282
80 282
81 282
82 282
83 282
84 282
85 282
86 282
87 282
88 282
89 282
90 282
91 282
92 282
93 282
94 282
95 282
96 282
97 282
98 282
99 282
100 282
101 282
102 282
103 282
104 282
105 282
106 282
107 282
108 282
109 282
110 282
111 282
112 282
113 282
114 282
115 282
116 282
117 282
118 282
119 282
120 282
};

\addplot [QiskitStyle]
table {
0 297.6
1 297.35
2 296.5
3 295.75
4 295.0
5 294.4
6 294.0
7 293.05
8 291.95
9 291.5
10 290.75
11 290.15
12 289.7
13 289.55
14 289.3
15 288.7
16 288.6
17 288.15
18 288.0
19 287.8
20 287.6
21 287.2
22 287.15
23 286.95
24 286.95
25 286.65
26 286.4
27 286.05
28 285.8
29 285.5
30 284.95
31 284.6
32 284.65
33 284.2
34 284.35
35 284.45
36 284.65
37 284.4
38 284.4
39 284.25
40 284.1
41 284.35
42 284.2
43 284.2
44 284.05
45 284.0
46 283.95
47 283.85
48 283.95
49 283.85
50 283.65
51 283.75
52 283.85
53 283.9
54 283.7
55 283.65
56 283.95
57 283.5
58 283.55
59 283.6
60 283.65
61 283.6
62 283.6
63 283.55
64 283.3
65 283.45
66 283.6
67 283.25
68 283.4
69 283.05
70 283.1
71 282.9
72 283.3
73 283.3
74 283.0
75 283.05
76 282.9
77 282.85
78 282.85
79 282.8
80 282.5
81 282.65
82 282.5
83 282.55
84 282.55
85 282.6
86 282.55
87 282.6
88 282.55
89 282.65
90 282.55
91 282.5
92 282.55
93 282.5
94 282.6
95 282.45
96 282.95
97 282.6
98 282.65
99 282.5
100 282.5
101 282.45
102 282.6
103 282.55
104 282.7
105 282.65
106 282.4
107 282.55
108 282.55
109 282.9
110 282.75
111 282.55
112 282.55
113 282.6
114 282.7
115 282.7
116 282.85
117 282.55
118 282.7
119 282.6
120 282.5
};

\addplot [MCTSStyle]
table {
0 298
1 299
2 298
3 297
4 296
5 296
6 296
7 295
8 293
9 293
10 293
11 292
12 291
13 291
14 291
15 290
16 291
17 290
18 290
19 289
20 289
21 288
22 289
23 289
24 290
25 288
26 288
27 288
28 288
29 288
30 286
31 286
32 287
33 285
34 286
35 287
36 288
37 286
38 286
39 285
40 285
41 286
42 286
43 287
44 286
45 286
46 286
47 285
48 287
49 286
50 285
51 285
52 287
53 287
54 285
55 285
56 288
57 285
58 285
59 285
60 286
61 285
62 286
63 286
64 284
65 285
66 286
67 285
68 285
69 285
70 285
71 284
72 286
73 285
74 284
75 285
76 284
77 285
78 284
79 285
80 284
81 284
82 284
83 284
84 284
85 284
86 284
87 285
88 284
89 285
90 284
91 283
92 284
93 284
94 284
95 284
96 287
97 284
98 284
99 283
100 284
101 283
102 285
103 284
104 284
105 285
106 283
107 284
108 285
109 285
110 285
111 284
112 284
113 284
114 284
115 285
116 285
117 284
118 284
119 284
120 284
};

\end{axis}
\end{tikzpicture}
    }
    \caption{The convergence of the proposed GA. Here the QCT circuit is QFT\_16 obtained by Qiskit on a 4x4 grid AG. For each iteration, the minimum, average and maximum fitness values within the population are recorded.
    }
    \label{fig:ga_con}
\end{figure}

\paragraph{Genetic algorithm for gate commutations}
To demonstrate the effectiveness of GA, we compare SSR2 with SSR3. The circuit depths can be further reduced by 11.65\% on average through gate commutation rules.
In order to demonstrate the convergence properties of GA, we independently execute Alg.~\ref{alg:ga} to optimize the QCT circuit QFT\_16 obtained by Qiskit on a 4x4 grid AG, and record, for each iteration, the minimum, average and maximum fitness values found. For the sake of conviction, we remove the termination conditions and force GA to be iterated 120 times. As depicted in Fig.~\ref{fig:ga_con}, the results illustrate that stable convergence is achieved after several dozen iterations, highlighting the efficacy of GA.

\begin{table}[h]
\centering
\caption{Evaluation of the ANN prediction model, where t(X) is the runtime (s) of algorithm X. Since the time of GA is non-deterministic, for a fair comparison, the runtime here does not consider the time required by GA}  
    \begin{tabular}{c|c|c|c|c}
    \hline
        Name & SAT Calls & t(SSR3) & SAT Calls & t(SSR) \\ \hline
        4mod5-bdd\_287  & {952}   & {17.9}  & 24  & 6.79  \\ \hline
        rd53\_135  & {3702}    & {145.6}  & 262   &  26.13  \\ \hline
        ham7\_104  & {3891}  & {157}   &  227  & 20.04  \\ \hline
        rd53\_131  & {7014}   & {259.8}  &  369  & 52.31  \\ \hline
        rd53\_133  & {9573}   & {791}  &  335  &  42.12  \\ \hline
        ex2\_227  & {9891}  & {798}   &  431  &  66.89  \\ \hline
        majority\_239  & {9146}  & {400.6}  & 429  &  89.32  \\ \hline
        rd53\_130  & {19380}  & {1287}   &  777  & 146.5  \\ \hline
        f2\_232  & {22494}   & {1677}  &  857 & 131.82  \\ \hline
        rd53\_251 & {25493} & {1964} &  892 &  176.15 \\ \hline
        Summation & 111536  & 7497.9 &  4603 & 758.07 \\ \hline
    \end{tabular}
    \label{tab:ann}
\end{table}

\paragraph{ANN depth prediction}

We employ an ANN to estimate the optimal depth of each candidate CNOT subcircuit, which serves two purposes in our framework.
First, the predicted depth is used as an initial target for SAT-based subcircuit rewriting (Sec.~\ref{sec:disable}), thereby reducing the number of failed SAT invocations without affecting the completeness of the SAT search for a given subcircuit.
Second, the predicted depth is used during the subcircuit sweeping process to prioritize which subcircuits are selected for SAT rewriting in each iteration.

As shown in Tab.~\ref{tab:ann}, incorporating ANN significantly reduces the overall execution time of the algorithm (by about 89.89\%), mainly due to a substantial decrease in the number of SAT solver calls (up to  95.87\% on average).
Regarding solution quality, the use of ANN has almost no effect on the final circuit depth (see columns 5 and 6 in Tab.~\ref{tab:disable}).
It is important to explicitly state that while the SAT component itself remains exact and guarantees local optimality for individual subcircuits, the overall SSR framework operates as a heuristic optimization architecture. The minor discrepancies observed in the final circuit depth between the ANN-assisted and non-ANN versions do not stem from any limitation or loss of exactness in the SAT solver. Rather, they arise from the heuristic interactions among the GA-based commutation, the subcircuit sweeping phase, and the ANN-guided subcircuit selection. Because the ANN predictions influence the optimization scores and dictate the rewrite ordering, the ANN inherently alters the overall search trajectory of the framework. This global sensitivity to rewrite ordering fully accounts for the experimental variations.

\subsection{Effectiveness of SSR}
\label{subsec:ssr_comp}

Now we will evaluate the overall performance of the comprehensive implementation of the proposed SSR algorithm. Our experiments were conducted on various AGs, including Grid 5x4, Google Sycamore, IBM Rochester, and IBM Heron\footnote{https://quantum.ibm.com/services/resources}. 
For Grid 5x4, we evaluate circuits from RevLib, which are widely used benchmarks in reversible and quantum circuit synthesis, covering a range of circuit sizes (6–16 qubits) and algebraic structures.
For the other three AGs, where the qubit count is significantly larger, we use a combination of real-world quantum circuits (including QFT and quantum volume circuits), parameterized circuit families (such as TwoLocal and IQP), as well as randomly generated circuits. These benchmarks capture diverse structural characteristics, including different entanglement patterns, gate densities, and degrees of regularity.

For circuits mapped to Sycamore and Rochester, we select only the first 200 layers of gates, and for Heron, the first 100 layers are used.
This choice reflects practical constraints of current NISQ devices, where executable circuit depths are typically limited to tens or a few hundred layers due to noise and coherence considerations.
Accordingly, evaluating the initial portion of each circuit provides a realistic setting for assessing optimization effectiveness.
Note that during gate selection, SWAP gates are not decomposed into CNOT gates.
For each tested AG, we generate input QCT circuits by compiling all benchmark circuits using two QCT algorithms, SABRE and STOCHASTIC, both of which have been integrated into Qiskit. The detailed experimental results are shown in Tab.~\ref{tab:dif_qcts}. 
Because GA is random, SSR was executed 5 times for each set of data and the average value was finally taken.
It can be observed that, regardless of the QCT method used, our SSR algorithm can significantly reduce circuit depth across all considered AGs (up to 29.04\% in Grid 5x4 and 16.59\% on average).

\begin{table}[!ht]
    \centering
\caption{
Evaluation for SSR on various AGs and circuits generated by different QCT methods. The values of \#Qubit, \#2Q Gates, \#SWAP are reported as ranges (min–max) over each benchmark set.
$\text{imp.} = \frac{X_{\text{ori}}-X_{\text{opt}}}{X_{\text{ori}}}$, where $X_{\text{ori}}$ represents the original value before optimization and $X_{\text{opt}}$ represents the value after optimization.
RT(s): the running time in seconds. All of these data are the geometric means of the results taken on all input circuits.}
    \begin{tabular}{c|c|c|c|c|c|c|c}
    \hline
        AG & QCT appro. & \#Qubits & \#2Q Gates & \#SWAPs & Depth imp. & Gate imp. & RT(s)  \\ \hline
        \multirow{2}{*}{Grid 5x4} & SABRE & 6-16 & 55-516& 17-180 & 29.04\% &  13.26\% & 
         49.66  \\ 
        ~ & STOCHASTIC & 6-16 & 63-570 & 25-234 &  27.12\% & 11.28\% &  70.22  \\ \hline
        \multirow{2}{*}{Sycamore} & SABRE & 20-53 & 57-1082 & 19-547 &  12.94\% & -0.05
        \% &  252.14  \\ 
        ~ & STOCHASTIC & 20-53 & 50-1921 & 12-1497 &  6.49\% &  0.48\% & 593.05   \\ \hline
        \multirow{2}{*}{Rochester} & SABRE & 20-53 & 61-1430 & 23-927 & 14.91\% & 1.42\% &  487.16   \\ 
        ~ & STOCHASTIC & 20-53 & 66-1686 & 28-1407 &14.79\% & 1.46\% &  580.86 \\ \hline
        \multirow{2}{*}{Heron} & SABRE & 53-156 & 94-1456 & 37-1239 & 13.12\% &  0.2\% & 261.41   \\ 
        ~ & STOCHASTIC & 53-156 & 248-1575 & 139-1379 & 14.29\% &  0.35\% & 547.34  \\ \hline
    \end{tabular}

    \label{tab:dif_qcts}
\end{table}

We also compare our SSR algorithm with two other depth optimization algorithms, CBIR~\cite{xie2021mitigating} and Q-Synth~\cite{ShaikvdP2024cnotsynthesis}.
CBIR leverages a two-level bidirectional reordering algorithm based on the gate commutation rules while taking connectivity constraints into account. Although CBIR is not directly applicable to optimizing circuit depth, it can be adapted for this purpose by modifying the cost function in Eq.~(5) in \cite{xie2021mitigating}. 
Q-Synth performs CNOT synthesis under explicit layout constraints. In our experiments, we adopt its \textit{peephole optimization} module, which extracts CNOT subcircuits (slices) from the input circuit and optimizes the depth of each extracted CNOT slice individually.
Since CBIR does not offer open-source programs, we independently reproduced the relevant code.
Tab.~\ref{tab:dif_ops} shows the experimental results. The reported statistics are collected directly from the layout-aware circuits after QCT, ensuring that the number of qubits and SWAP operations accurately reflect the actual hardware-constrained implementations.
It can be observed that SSR significantly outperforms the other two methods in terms of the circuit depth.

While Q-Synth performs well on the Grid 5x4 architecture, we observe that on larger AGs it not only fails to reduce circuit depth but may even increase it. A likely reason is that Q-Synth was originally evaluated on relatively small circuits and its scalability to large-qubit settings has not been thoroughly explored. Moreover, its optimization operates on extracted CNOT slices in isolation, without explicit mechanisms to balance local improvements against global circuit structure. In large-scale circuits, such purely local rewritings may inadvertently introduce unfavorable interactions across subcircuits, resulting in increased overall depth.

Although the goal of SSR is to optimize depth, experimental results show that SSR can still reduce the number of gates while optimizing depth. As summarized in Tab.~\ref{tab:dif_ops}, SSR can achieve non-negligible gate-count improvements on most architectures (13.26\% at most on Grid 5x4).

\begin{table}[!ht]
    \centering
    \caption{Comparison of different post-QCT optimization methods on SABRE-mapped circuits. Depth imp. and gate imp. have the same meaning as Tab.~\ref{tab:dif_qcts}. }
    \begin{tabular}{c|c|c|c}
    \hline
        AG & Method & Depth imp. & Gate imp.   \\ \hline
        \multirow{4}{*}{Grid 5x4} & SSR & 29.04\% &  13.26\%  \\ 
        ~ & CBIR & 12.17\% & 0.00\%  \\ 
         ~ & Q-Synth & 16.63\% & 5.96\%\\ \hline
        \multirow{4}{*}{Sycamore} & SSR &  13.36\% & -0.05\%  \\ 
        ~ & CBIR & 5.80\% & 0.00\%  \\ 
        ~ & Q-Synth & -5.88\% & -11.13\% \\ \hline
        \multirow{4}{*}{Rochester} & SSR &  14.91\% & 1.42\%  \\ 
        ~ & CBIR & 3.71\% & 0.00\%  \\ 
        ~ & Q-Synth & -10.53\% & -16.3\% \\ \hline 
        \multirow{4}{*}{Heron} & SSR &  13.12\% & 0.2\%  \\ 
        ~ & CBIR & 4.11\% & 0.00\%  \\ 
        ~ & Q-Synth & -2.33\% & -3.28\% \\ \hline
    \end{tabular}
    
    \label{tab:dif_ops}
\end{table}

To further examine the robustness of SSR, we additionally evaluate its performance on circuits that have been pre-optimized using Qiskit. The same set of original (unmapped) benchmark circuits as those used in the previous experiments is adopted here to ensure a fair and consistent comparison. Qiskit offers several preset optimization passes, controlled by the parameter \emph{optimization\_level}, which ranges from 0 to 3\footnote{https://quantum.cloud.ibm.com/docs/en/api/qiskit/qiskit.transpiler.generate\_preset\_pass\_manager}. For all \emph{optimization\_level} values, the SABRE algorithm is applied to ensure that the resulting circuits comply with device connectivity constraints, while additional gate-level simplification techniques, e.g., gate cancellation and KAK-based 2-qubit decomposition, will be incorporated as \emph{optimization\_level} increases. These pre-optimized circuits are then passed to SSR to further reduce their depths. As reported in Tab.~\ref{tab:com_qiskit}, the depth improvement achieved by SSR gradually decreases as the Qiskit optimization level rises, which is expected since a higher \emph{optimization\_level} will eliminate more available optimization space.

Nevertheless, even when applied after Qiskit optimization at level~3, SSR still achieves around 10\% depth reduction on average. Moreover, this depth improvement is obtained with only a marginal increase in gate count, indicating that SSR remains effective as a post-optimization technique and complements existing compiler-level optimizations.

\begin{table}[!ht]
    \centering
    \caption{Performance of SSR for circuits optimized by different Qiskit optimization levels. The Depth imp. and Gate imp. values are computed relative to circuits produced at the respective Qiskit optimization levels.}
    \label{tab:com_qiskit}
    \begin{tabular}{c|c|c|c}
    \hline
        AG & Opt. Level & Depth imp. & Gate imp.   \\ \hline
        \multirow{4}{*}{Grid 5x4} & 0 &  30.03\% &  9.85\%  \\ 
        ~ & 1 & 28.37\% &  12.98\%  \\ 
        ~ & 2 & 15.55\% & 4.19\%  \\ 
         ~ & 3 &  15.42\% & 3.69\% \\ \hline
        \multirow{4}{*}{Sycamore} & 0 &  12.21\% & 0.00\%  \\ 
        ~ & 1 & 12.99\% & 0.17\%  \\ 
        ~ & 2 &  10.27\% & -0.65\% \\ 
        ~ & 3 & 10.15\% &  -0.59\% \\ \hline
        \multirow{4}{*}{Rochester} & 0 & 12.89\% & 0.93\%  \\ 
        ~ & 1 & 12.92\% & 0.39\%  \\ 
        ~ & 2 & 10.66\% & -0.2\% \\ 
        ~ & 3 &  10.29\% & -0.00\% \\ \hline 
        \multirow{4}{*}{Heron} & 0 &  11.1\% & 0.25\%  \\ 
        ~ & 1 &  10.21\% & -0.08\%  \\ 
        ~ & 2 &  10.56\% &  -0.43\%  \\
        ~ & 3 & 9.45\% & -0.39\%\\ \hline
        \end{tabular}
    
\end{table}

\begin{figure}
\centering
\pgfplotsset{
        index_axis/.style={
            xlabel={Qubit number},
            ylabel={Runtime(s)},
            xtick={20, 30, 40, 50, 60, 70, 80, 90, 100},
            xticklabel style={
                /pgf/number format/fixed,
                /pgf/number format/precision=3,
                rotate=45,
                anchor=east
            },
            width = 12cm,
            height = 8cm,
            grid=both,
            mark repeat=1,
            xlabel style = {font =\Large},
            ylabel style = {font =\Large},        
        },
        gates_axis/.style={
            xlabel={Average gate number},
            ylabel={Runtime(s)},
            xtick={600, 800, 1000, 1200, 1400, 1600, 1800},
            xticklabel style={
                /pgf/number format/fixed,
                /pgf/number format/precision=3,
                rotate=45,
                anchor=east
            },
            width = 12cm,
            height = 8cm,
            grid=both,
            mark repeat=1,
            xlabel style = {font =\Large},
            ylabel style = {font =\Large},
        }
        }
\subfloat[Runtime vs. number of qubits]{
    
    \scalebox{0.57}{
    \begin{tikzpicture}
\begin{axis}[index_axis ]

\addplot [SSRStyle]
table {%
20	52.2482621
25	67.76047672
36	108.4758781
49	134.0642177
64	175.1559041
81	192.1997833
100	272.1008824
};

\end{axis}
\end{tikzpicture}
    }
    }
\hfill
\subfloat[Runtime vs. the average number of gates, where the average is taken over 10 random circuits for each AG.]{
    \scalebox{0.57}{
    \begin{tikzpicture}
\begin{axis}[gates_axis ]

\addplot [SSRStyle]
table {%
602.9	52.2482621
680.7	67.76047672
884.3	108.4758781
1078.6	134.0642177
1291.4	175.1559041
1552	192.1997833
1776.3	272.1008824
};

\end{axis}
\end{tikzpicture}
    }
}   
\caption{Scalability of SSR on random circuits with depth 200 over grid architectures of different sizes (from 5$\times$4 to 10$\times$10).  For each AG, 10 random circuits are tested, and the average runtime is reported. }
\label{fig:scale}
\end{figure}

To evaluate SSR's scalability in terms of the qubit number, we conduct experiments on grid AGs from 5x4 to 10x10, corresponding to 20, 25, 36, 49, 64, 81, and 100 qubits. For each AG, 10 random circuits {with a fixed depth of 200} are tested, and the corresponding QCT circuits are obtained by SABRE. As shown in Fig.~\ref{fig:scale}(a), the running time grows almost linearly with the number of qubits, demonstrating that SSR is scalable in terms of qubit numbers. 
Fig.~\ref{fig:scale}(b) plots the average runtime against the average number of gates of the 10 random circuits for each qubit size. A similar near-linear growth trend is observed, indicating that SSR is also scalable with respect to the circuit size.
Although our SSR method has a longer runtime compared to purely heuristic algorithms, it can be further optimized by implementing it with other compiled programming languages or utilizing parallel computing techniques.

\section{Conclusion}

In this work, we present the SSR optimizer to further reduce the depth of quantum circuits outputted by the QCT process,  while keeping the circuits compliant with the underlying physical constraints. By integrating the SAT Sweeping technique, the gate commutation rule, and the ANN prediction model, the proposed method has achieved a remarkable effect in reducing circuit depth. The experimental results demonstrate that our proposed method can achieve universal optimization effects on quantum circuits outputted by various QCT methods across different architectures.

\section{Acknowledgments}

Work partially supported by the Innovation Program for Quantum Science and Technology (No. 2021ZD0302901), the Young Scientists Fund of the Natural Science Foundation of Jiangsu Province (No. BK20240536), the Jiangsu Funding Program for Excellent Postdoctoral Talent (No.2022ZB139), the Jiangsu Province Natural Science Foundation of China under Grant (No. BK20221411), the National Natural Science Foundation of China (No. 62502203), and the Natural Science Foundation of Jiangsu Province Higher Education Basic Science (No. 24KJB120005).

\section*{Ethics and Privacy Statement}

This section of your ACM work should discuss the potential societal
risks that might result from its publication; two to three sentences
related to the findings of your study, or new advancements made
possible by their developed methods. The privacy and ethics statement
should clearly address the broader impacts of their work as it relates
to the authors' interpretation of privacy, fairness, safety, human
rights, data sovereignty, or future misuse and any benefit/risk
trade-off resulting from this research. We acknowledge that some
papers may have minimal societal risks beyond those considered by
institutional review boards, and the dimensions considered by any
review of the user study design or dataset licenses could be provided
in this statement.

\bibliographystyle{ACM-Reference-Format}
\bibliography{ref}


\end{document}